\documentclass[11pt,a4paper]{article}
\pdfoutput=1

\usepackage{amsmath}
\usepackage[T1]{fontenc}

\usepackage{jheppub}
\usepackage{psfrag}
\usepackage{slashed}
\usepackage{cancel}
\usepackage{lscape}

\usepackage{caption}
\usepackage{array}
\usepackage{graphicx}
\usepackage{subcaption}
\usepackage{multirow}
\usepackage{tabularx}
\usepackage{makecell}
\usepackage{hyperref}

\usepackage[utf8]{inputenc}

\def\cF{\mathcal{F}}

\def\cH{\mathcal{H}}
\def\cA{\mathcal{A}}

\def\nn{\nonumber \\ }

\def\vareps{\varepsilon}

\def\la{\langle}
\def\ra{\rangle}
\def\spA#1#2{\la#1#2\ra}
\def\spB#1#2{[#1#2]}

\DeclareMathOperator{\tr}{\rm tr}
\def\trm{\tr_-}
\def\trp{\tr_+}

\def\eps{\epsilon}

\def\cusp{{\mathrm{cusp}}}

\def\wpaj{W^+\gamma j}
\def\wmaj{W^-\gamma j}

\newcolumntype{C}[1]{>{\hsize=#1\hsize\centering\arraybackslash}X}%

\newcolumntype{Z}{r<{\hspace{3mm}}}

\newcommand{\ba}{\[\begin{aligned}}
\newcommand{\ea}{\end{aligned}\]}

\title{Two-loop leading colour helicity amplitudes for $W^\pm\gamma+j$ production at the LHC}

\author[a]{Simon Badger,}
\author[b]{Heribertus Bayu Hartanto,}
\author[a,c]{Jakub Kry\'s,}
\author[a]{Simone Zoia}

\affiliation[a]{
Dipartimento di Fisica and Arnold-Regge Center, Università di Torino, and INFN, Sezione di
Torino, Via P. Giuria 1, I-10125 Torino, Italy
}

\affiliation[b]{
Cavendish Laboratory, University of Cambridge, Cambridge CB3 0HE, United Kingdom
}

\affiliation[c]{
Institute for Particle Physics Phenomenology, Department of Physics, Durham University, Durham DH1 3LE, United Kingdom
}

\emailAdd{
simondavid.badger@unito,it,
hbhartanto@hep.phy.cam.ac.uk,
jakubmarcin.krys@unito.it,
simone.zoia@unito.it
}

\abstract{
We present the two-loop leading colour QCD helicity amplitudes for the process $pp\to W(\to
l\nu)\gamma+j$. We implement a complete reduction of the amplitudes, including the leptonic decay of the $W$-boson,
using finite field arithmetic, and extract the analytic finite remainders using a recently
identified basis of special functions. Simplified analytic expressions are obtained after considering
permutations of a rational kinematic parametrisation and multivariate partial fractioning. We
demonstrate efficient numerical evaluation of the two-loop colour and helicity summed finite
remainders for physical kinematics, and hence the suitability for phenomenological applications.
}

\preprint{CAVENDISH-HEP-22/01}

\begin{document}
\maketitle
\flushbottom
\section{Introduction \label{sec:intro}}

High-precision theoretical predictions are a high priority for the current experiments at the Large
Hadron Collider (LHC). Processes with a pair of electroweak vector bosons $(W^\pm,Z,\gamma)$ offer a wide
range of observables which can test electroweak couplings and probe the Higgs sector of the Standard
Model. In particular, the production of a $W$ boson in association with a photon ($pp\to W\gamma$) is
one of the processes observed at the LHC with relatively large cross sections where clean
signatures can be acquired when the $W$ boson decays leptonically. $W\gamma$ production enables direct access to
the $WW\gamma$ triple gauge boson coupling, which can be modified in certain new physics scenarios.
Both ATLAS and CMS experiments have measured the $W\gamma$ process~\cite{CMS:2011myh,ATLAS:2011nmx,ATLAS:2012bpb,ATLAS:2013way,CMS:2013ryd,CMS:2021foa} and set the limit on the
anomalous $WW\gamma$ coupling.

Predictions for $pp\to W^\pm\gamma$ are available through to next-to-next-to-leading order (NNLO) in
QCD~\cite{Gehrmann:2011ab,Grazzini:2015nwa,Campbell:2021mlr} and NLO in the electroweak (EW)
coupling~\cite{Accomando:2005ra,Denner:2014bna} as well as combined NNLO in QCD and NLO in EW~\cite{Grazzini:2019jkl}. The colourless final state makes the process well
suited for the $q_T$~\cite{Catani:2007vq} and $N$-jettiness~\cite{Boughezal:2015aha,Gaunt:2015pea} subtraction methods as implemented within the \textsc{Matrix}~\cite{Grazzini:2017mhc}
and \textsc{MCFM}~\cite{Boughezal:2016wmq} Monte Carlo event generators, respectively. Resummed predictions including parton shower
effects are now available~\cite{Cridge:2021hfr}, making this one of the most precisely known
theoretical predictions. Experimental measurements are constantly improving and provide rich
opportunities for precision SM tests~\cite{CMS:2021foa,CMS:2021cxr}.
In order to suppress different types of backgrounds in the experimental analysis, it is a common
practice to divide the measurement according to the jet multiplicities, i.e.\  $W\gamma$+0 jet,
$W\gamma$+1 jet, $W\gamma$+2 jets, etc. Increasing the precision of the theoretical predictions for
each of the jet bins amounts to computing higher order corrections to $W\gamma$ production in association
with additional jets.

In this article we compute the two-loop helicity amplitudes for the process $pp\to W^\pm(\to l^\pm\nu)\gamma + j$ for
the first time. The amplitude-level ingredients we provide will give useful information for future precision 
measurements of anomalous couplings and potentially for complete global fits of the Standard Model
Effective Theory (SMEFT). A fully differential computation of $W\gamma+j$ at NNLO in QCD would also
open up the possibility of N${}^3$LO QCD predictions for $W\gamma$ production.

The process has already been well studied at NLO in QCD~\cite{Campanario:2009um}, including the anomalous couplings~\cite{Campanario:2010hv}, 
and is easily within the reach of standard automated tools including electroweak corrections. Compact analytic formulae at one loop have been obtained recently~\cite{Campbell:2021mlr}.

Recent years have seen tremendous progress in the analytic computation of two-loop amplitudes for processes involving five massless particles~\cite{Gehrmann:2015bfy,Badger:2018enw,Abreu:2018aqd,Chicherin:2018yne,Chicherin:2019xeg,Abreu:2019rpt,%
Abreu:2018zmy,Abreu:2018jgq,Abreu:2019odu,Badger:2019djh,Abreu:2020cwb,Chawdhry:2020for,Caron-Huot:2020vlo,Chicherin:2020oor,DeLaurentis:2020qle,Agarwal:2021grm,Abreu:2021oya,%
Chawdhry:2021mkw,Agarwal:2021vdh,Badger:2021imn}.
The methods are now reaching maturity also for two-loop five-particle amplitudes with an external off-shell leg, with the
first sets of analytic helicity amplitudes in the planar limit appearing over the last 12
months~\cite{Badger:2021nhg,Badger:2021ega,Abreu:2021asb}. This progress has been made possible by the classification of the special functions appearing in the finite remainders through the
differential equations they satisfy~\cite{Papadopoulos:2015jft,Papadopoulos:2019iam,Abreu:2021smk,Abreu:2020jxa,
Canko:2020ylt,Syrrakos:2020kba,Chicherin:2021dyp}. Furthermore, finite-field arithmetic for
scattering amplitudes~\cite{vonManteuffel:2014ixa,Peraro:2016wsq,Peraro:2019svx} allows us to avoid the large intermediate expressions and provides efficient solutions to large systems of
integration-by-parts integral identities~\cite{Tkachov:1981wb,Chetyrkin:1981qh,Laporta:2000dsw}, which are the main bottlenecks in the analytic computation of multi-loop scattering amplitudes.

Our paper is organised as follows. In Section~\ref{sec:amp} we describe the structure of the
amplitudes for $pp\to W\gamma j$ up to two loops, paying particular attention to the description of
the decay of the $W$ boson. In Section~\ref{sec:reduction} we describe the finite field reduction
setup used to extract the finite remainders, and propose an approach to simplify dramatically the analytic expressions of the latter based on a systematic search of a better parameterisation of the kinematics in terms of momentum-twistor variables. We describe a number of validation tests that have been
performed on our results in Section~\ref{sec:validation}, and then present some numerical results for the
colour and helicity summed finite remainders in Section~\ref{sec:results}. We
present our conclusions and outlook for the future in Section~\ref{sec:conclusions}. Complete analytic expressions are provided in
the associated ancillary files on the \texttt{arXiv}. In addition, we include appendices describing the details
of the renormalisation constants, and giving explicit one-loop results to facilitate future cross-checks.

\section{Structure of the Amplitudes}
\label{sec:amp}

\begin{figure}[t!]
  \begin{center}
    \includegraphics[width=0.75\textwidth]{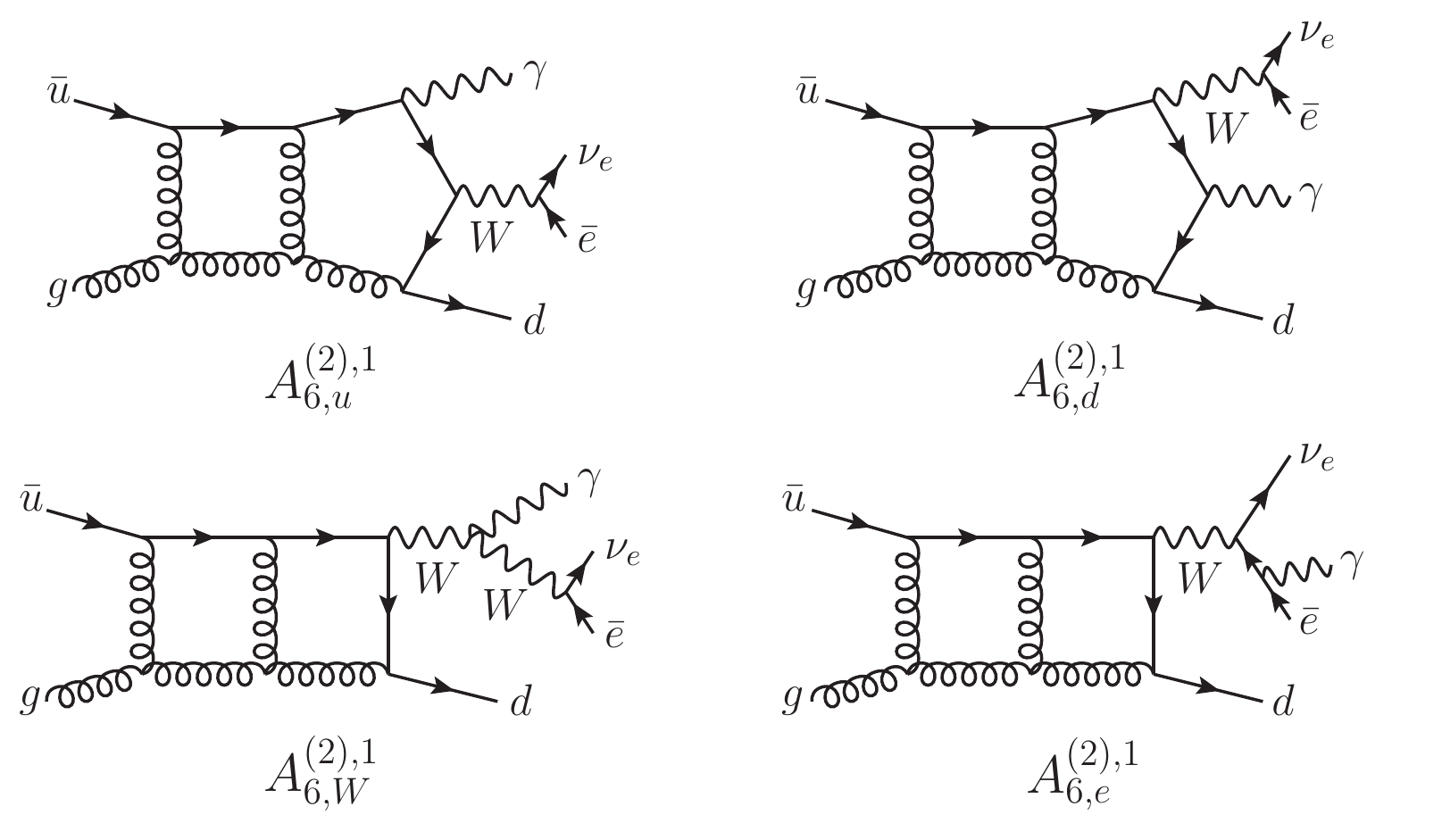}
  \end{center}
  \caption{Sample two-loop Feynman diagrams for $\wpaj$ production.}
  \label{fig:diag2L}
\end{figure}

\begin{figure}[t]
  \begin{center}
    \includegraphics[width=0.72\textwidth]{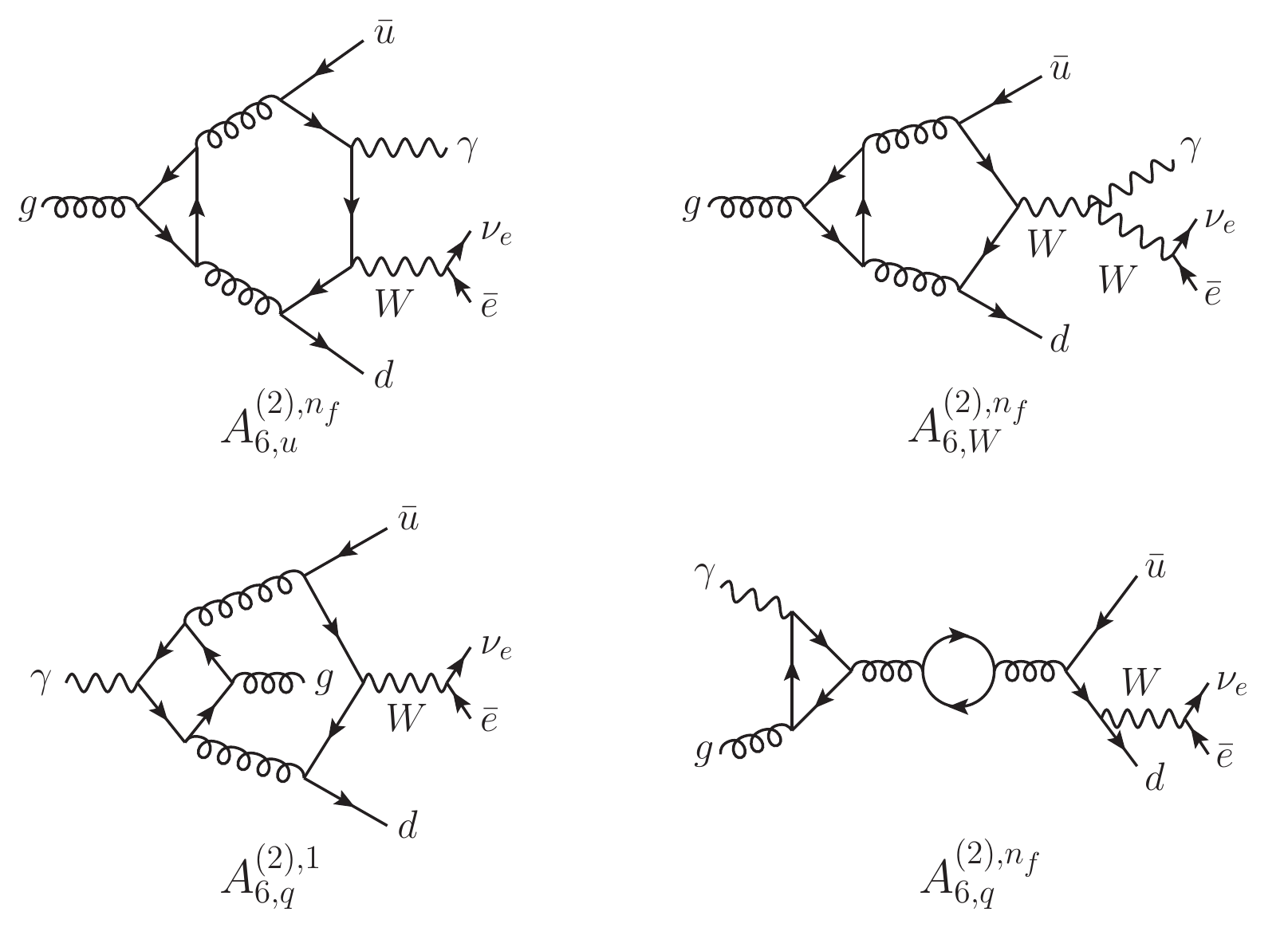}
  \end{center}
  \caption{Sample two-loop Feynman diagrams for $\wpaj$ production containing closed fermion loop. $A_{6,q}^{(2),n_f}$ vanishes due to Furry's theorem.}
  \label{fig:diag2Lnf}
\end{figure}

We compute the two-loop amplitudes for the production of a $W^+$ boson in association with a photon and a jet at hadron colliders,
where the $W^+$ boson decays to a positron and an electron neutrino ($pp \to \nu_e e^+ \gamma j$),
in the leading colour approximation,
\begin{equation} \label{eq:wpaj}
0 \rightarrow \gamma(p_1,h_1)+\bar{u}(p_2,h_2)+g(p_3,h_3)+d(p_4,h_4)+\nu_e(p_5,h_5)+e^+(p_6,h_6) \,.
\end{equation}
For simplicity we denote this process as $\wpaj$ production henceforth. Sample two-loop Feynman diagrams contributing at leading colour are shown in Figures~\ref{fig:diag2L} and~\ref{fig:diag2Lnf}.
The colour decomposition of the $\wpaj$ $L$-loop amplitude is given by
\begin{equation}
\cA_{6}^{(L)}(1_\gamma,2_{\bar{u}},3_g,4_d,5_{\nu},6_{\bar{e}}) = \sqrt{2} e g_W^2  g_s \, n^L \, (T^{a_3})_{i_4}^{\;\;\bar i_2} \, A_{6}^{(L)}(1_\gamma,2_{\bar{u}},3_g,4_d,5_{\nu},6_{\bar{e}}) \,,
\label{eq:colourdecomposition}
\end{equation}
where  $n= m_\eps  \alpha_s/(4\pi)$, $\alpha_s = g_s^2/(4\pi)$, $m_\eps=i (4\pi)^{\eps} e^{-\eps\gamma_E}$, $\eps = (4-d)/2$ is the dimensional regulator, and
$T^a$ are the generators of $SU(N_c)$ in the fundamental representation, normalised according to $\tr(T^a T^b) = \delta^{ab}$.
We denote by $e$, $g_W$ and $g_s$ the electron charge, the weak and the strong coupling constants, respectively.
The $\wpaj$ amplitude can be further decomposed according to the source of photon radiation,
\begin{align}
\label{eq:ampdecomposition}
\begin{aligned}
A_{6}^{(L)} = \ & \left[ Q_u A^{(L)}_{6,u} + Q_d A^{(L)}_{6,d} + \left(\sum_{q} Q_q\right) A^{(L)}_{6,q} \right] P(s_{56}) \\
& + \left(Q_u - Q_d\right) \left[ A^{(L)}_{6,e} + A^{(L)}_{6,W} P(s_{56}) \right] P(s_{156}) \,,
\end{aligned}
\end{align}
where $s_{ij}=(p_i+p_j)^2$ and $s_{ijk}=(p_i+p_j+p_k)^2$, $Q_u$ and $Q_d$ are the up- and down-quark charges respectively, the sum runs over the quark flavours $q$, while 
\begin{equation}
P(s) = \frac{1}{s-M_W^2+i M_W \Gamma_W} \,,
\end{equation}
is the denominator factor of $W$ boson propagator. 
$M_W$ and $\Gamma_W$ are the mass and decay width of the $W$ boson, respectively.
The sub-amplitudes $A_{6,i}^{(L)}$ in Eq.~\eqref{eq:ampdecomposition} are categorised as follows:
\begin{itemize}
\item $A^{(L)}_{6,u}$: the photon is radiated off the $u$ quark;
\item $A^{(L)}_{6,d}$: the photon is radiated off the $d$ quark;
\item $A^{(L)}_{6,W}$: the photon is radiated off the $W$ boson;
\item $A^{(L)}_{6,e}$: the photon is radiated off the positron;
\item $A^{(L)}_{6,q}$: the photon is radiated off the internal quark loop.
\end{itemize}
We stress that the sub-amplitudes are not separately gauge invariant in the electroweak sector.
Using the relation~\cite{Campbell:2021mlr}
\begin{equation}
P(s_{56})P(s_{156}) = \frac{1}{s_{156}-s_{56}} \bigg[ P(s_{56}) - P(s_{156}) \bigg] 
\end{equation}
we can rewrite Eq.~\eqref{eq:ampdecomposition} as
\begin{align}
A_{6}^{(L)} & =   Q_u \bigg[ A^{(L)}_{6,u} + \frac{1}{s_{156}-s_{56}} A^{(L)}_{6,W} \bigg] P(s_{56})
                + Q_d \bigg[ A^{(L)}_{6,d} - \frac{1}{s_{156}-s_{56}} A^{(L)}_{6,W} \bigg] P(s_{56}) \nonumber \\
&         \quad + (Q_u - Q_d) \bigg[ A^{(L)}_{6,e} - \frac{1}{s_{156}-s_{56}} A^{(L)}_{6,W} \bigg] P(s_{156})
          + \left(\sum_q Q_q\right) A^{(L)}_{6,q} P(s_{56}) \,, 
\label{eq:gaugeinvariantamps}
\end{align}
such that the combinations of sub-amplitudes in the square brackets and $A^{(L)}_{6,q}$ are the gauge invariant pieces.
We further decompose the sub-amplitudes according to their closed fermion loop contributions. At leading colour we have
\begin{align} \label{eq:NFdecomposition}
\begin{aligned}
& A^{(1)}_{6,i} = N_c A^{(1),1}_{6,i} \,, \\
& A^{(2)}_{6,i} = N_c^2 A^{(2),1}_{6,i} + N_c n_f A^{(2),n_f}_{6,i} \,, \\
& A^{(2)}_{6,q} = N_c A^{(2),1}_{6,q}  \,,
\end{aligned}
\end{align}
where  $i=u,d,W,e$, and $n_f$ is the number of massless quark flavours running in the loop. 
We note that $A^{(L)}_{6,q}$ vanishes at tree level and one loop, while at two loops it includes non-planar contributions, and thus will not be considered in this work.

The coupling of the $W$ boson to fermions involves vector and axial-vector ($V-A$) vertices in the form of $\gamma^\mu(1-\gamma_5)/2$.
The massless fermion pairs that are coupled to the $W$ boson are connected to the external states and the $V-A$ coupling fixes the helicity of the fermion pairs. Therefore we only need to take into account the vector coupling of the $W$ boson to fermions when computing the helicity amplitudes. The contributing helicity configurations are 
\begin{equation}
A_{6}^{(L)}(1^\pm_\gamma,2^+_{\bar{u}},3^\pm_g,4^-_d,5^-_{\nu},6^+_{\bar{e}}) \,. 
\nn
\end{equation}
We choose $\scriptstyle +++--+$ and $\scriptstyle -++--+$ as the independent helicity configurations and focus on them. We obtain the amplitudes in the other helicity configurations from the independent ones by suitably permuting the external momenta and conjugating space-time parity.

The sub-amplitudes $A^{(L)}_{6,u}$ and $A^{(L)}_{6,d}$ are related by
\begin{equation}  \label{eq:ufromd}
A_{6,u}^{(L)}(1^{h_1}_\gamma,2^+_{\bar{u}},3^{h_3}_g,4^-_d,5^{h_5}_{\nu},6^{h_6}_{\bar{e}})
=
A_{6,d}^{(L)}(1^{h_1}_\gamma,4^-_{\bar{u}},3^{h_3}_g,2^+_d,5^{h_5}_{\nu},6^{h_6}_{\bar{e}}) \,.
\end{equation}
As a result, we can limit ourselves to computing the $A^{(L)}_{6,d}$ sub-amplitudes with the following helicity configurations,
\begin{equation}
A_{6,d}^{(L)}(1^\pm_\gamma,2^\pm_{\bar{u}},3^\pm_g,4^\mp_d,5^-_{\nu},6^+_{\bar{e}})\,.
\nn
\end{equation}
The independent helicity amplitudes for $A^{(L)}_{6,u}$ are then obtained through
\begin{equation}
A_{6,u}^{(L)}(1^\pm_\gamma,2^+_{\bar{u}},3^+_g,4^-_d,5^-_{\nu},6^+_{\bar{e}})
=
A_{6,d}^{(L)}(1^\pm_\gamma,4^-_{\bar{u}},3^+_g,2^+_d,5^-_{\nu},6^+_{\bar{e}}) \,.
\end{equation}

The pole structure of the unrenormalised $\wpaj$ amplitudes in the 't Hooft-Veltman (tHV) scheme at one and two loops is given by~\cite{Catani:1998bh,Becher:2009qa,Becher:2009cu,Gardi:2009qi}
\begin{align}
P_6^{(1)} & = 2 I_1(\eps) + \frac{\beta_0}{2\eps},
\label{eq:poles1L} \\
P_6^{(2)} & =   2 I_1(\eps) \bigg( \hat{A}^{(1)}_6 - \frac{\beta_0}{2\eps} \bigg) + 4 I_2(\eps)
                +  \frac{3\beta_0}{2\eps}  \hat{A}^{(1)}_6  - \frac{3\beta_0^2}{8\eps^2} + \frac{\beta_1}{4\eps}  \,,
\label{eq:poles2L}
\end{align}
where $\hat{A}^{(1)}_6$ is the unrenormalised one-loop amplitude divided by the tree-level amplitude. 
The $I_2(\eps)$ operator is given by
\begin{equation}
I_2(\eps) =   - \frac{1}{2}I_1(\eps) \left[ I_1(\eps)
              + \frac{\beta_0}{\eps} \right]
              + \frac{N(\eps)}{N(2\eps)} \left[ \frac{\beta_0}{2\eps}
                        + \frac{\gamma_{1}^{\cusp}}{8} \right] I_1(2\eps)
              + H^{(2)}(\eps) \,,
\end{equation}
while the $I_1(\eps)$ operator is given at leading colour by
\begin{align}
I_1(\eps) &= -N_c \frac{N(\eps)}{2} \bigg( \frac{1}{\eps^2} + \frac{3}{4\eps} + \frac{\beta_0}{4 N_c \eps} \bigg) \big[ \left(-s_{23}\right)^{-\eps} + \left(-s_{34}\right)^{-\eps} \big] \,,
\label{eq:I1}
\end{align}
where $N(\eps) = {e^{\eps \gamma_E}}/{\Gamma(1-\eps)}$ and
\begin{align}
H^{(2)}(\eps) &= \frac{1}{16\eps} \bigg\lbrace \big(2\gamma_1^q+\gamma_1^g\big)
                                                - \gamma_1^\cusp \left(\frac{\gamma_0^q}{2}+\frac{\gamma_0^g}{4}\right)
                                                + \frac{\pi^2}{8} \beta_0 \gamma_0^\cusp \left(C_F+\frac{C_A}{2}\right) \bigg\rbrace \,.
\end{align}
The $\beta$ function coefficients and anomalous dimensions are tabulated in Appendix~\ref{app:renormconstants}. We stress that the pole terms in Eqs.~\eqref{eq:poles1L} and~\eqref{eq:poles2L} include both the ultraviolet and infrared singularities.
We then extract the $L$-loop partial finite remainder by subtracting the poles $P_6^{(L)}$ from the unrenormalised partial amplitude $A_6^{(L)}$ and sending $\eps$ to $0$,
\begin{equation}
F_6^{(L)} = \lim_{\eps \to 0} \left[ A_6^{(L)} - P_6^{(L)} A_6^{(0)} \right]\,.
\label{eq:finiteremainder}
\end{equation}
The finite remainder $F_6^{(L)}$ follows the same decomposition as the unrenormalised partial amplitude $A_6^{(L)}$ (see Eqs.~\eqref{eq:ampdecomposition} and~\eqref{eq:NFdecomposition}),
\begin{equation}
\begin{aligned}
F_{6}^{(L)} = \,  \left[ Q_u F^{(L)}_{6,u} + Q_d F^{(L)}_{6,d}  \right] P(s_{56})  + (Q_u - Q_d) \left[ F^{(L)}_{6,e} + F^{(L)}_{6,W} P(s_{56}) \right] P(s_{156}) \,,
\end{aligned}
\label{eq:finremdecomposition}
\end{equation}
with
\begin{align} \label{eq:NfDecompositionF} 
\begin{aligned}
& F^{(1)}_{6,i} = N_c F^{(1),1}_{6,i} + n_f F^{(1),n_f}_{6,i} \,, \\
& F^{(2)}_{6,i} = N_c^2 F^{(2),1}_{6,i} + N_c n_f F^{(2),n_f}_{6,i} + n_f^2 F^{(2),n_f^2}_{6,i} \,,
\end{aligned}
\end{align}
where  $i=u,d,W,e$. We note that, although the bare sub-amplitudes $A^{(1),n_f}_{6,i}$ and $A^{(2),n_f^2}_{6,i}$ vanish, there are finite contributions to the finite remainders $F^{(1),n_f}_{6,i}$
and $F^{(2),n_f^2}_{6,i}$ from the UV renormalisation and IR subtraction terms specified in Eqs.~\eqref{eq:poles1L}~and~\eqref{eq:poles2L}. 
As discussed below Eqs.~\eqref{eq:NFdecomposition}, we defer the computation of $F^{(2)}_{6,q}$, as it involves the non-planar integrals.

For the charge-conjugated process, i.e.\ $pp \to \bar\nu_e e^- \gamma j$, we consider the amplitudes for
\begin{equation}
0 \rightarrow \gamma(p_1,h_1)+\bar{d}(p_2,h_2)+g(p_3,h_3)+u(p_4,h_4)+e^-(p_5,h_5)+\bar\nu_e(p_6,h_6) \,, \nonumber
\end{equation}
which we denote by $\wmaj$ production. The amplitudes for $\wmaj$ production can be obtained from the $\wpaj$ results through the following relation,
\begin{equation}  \label{eq:wmajamps}
A_{6}^{(L)}(1^{-h_1}_\gamma,2^{+}_{\bar{d}},3^{-h_3}_g,4^{-}_u,5^{-}_{e},6^{+}_{\bar\nu}) =  
\bigg[ A_{6}^{(L)}(1^{h_1}_\gamma,4^{+}_{\bar{u}},3^{h_3}_g,2^{-}_d,6^{-}_{\nu},5^{+}_{\bar{e}})\bigg]_{\spA{i}{j} \leftrightarrow \spB{i}{j}} \,.
\end{equation}

\section{Amplitude Computation}
\label{sec:reduction}
In this section we describe the computation of the two-loop $\wpaj$ amplitude in the leading colour approximation.
To derive the analytic form of the amplitude we employ a framework that combines Feynman diagram input, the four-dimensional projector method~\cite{Peraro:2019cjj,Peraro:2020sfm}, integration-by-parts (IBP) reduction~\cite{Tkachov:1981wb,Chetyrkin:1981qh,Laporta:2000dsw},
Laurent expansion onto a basis of special function, numerical evaluations over finite fields, and analytic reconstruction techniques~\cite{vonManteuffel:2014ixa,Peraro:2016wsq,Peraro:2019svx}.

Instead of computing the loop amplitudes using the full six-particle kinematics, 
we detach the leptonic decay of the $W$ boson from the amplitude and only compute the $W$-production amplitudes.
For the $A^{(L)}_{6,u}$ and $A^{(L)}_{6,d}$ sub-amplitudes the $W$-production amplitude is a five-point amplitude with an off-shell leg (denoted by  $A_{5,u/d}^{^{(L)}\mu}$),
while for the $A^{(L)}_{6,W}$ and $A^{(L)}_{6,e}$ sub-amplitudes the $W$-production amplitude is a four-point amplitude with an off-shell leg (denoted by $A_{4}^{{(L)}\mu}$),
\begin{align}
& A^{(L)}_{6,u/d}(p_1,p_2,p_3,p_4,p_5,p_6) = A_{5,u/d}^{(L)\mu}(p_1,p_2,p_3,p_4,p_W) \; L_{A,\mu}(p_5,p_6) \,, \label{eq:decayA}\\
& A^{(L)}_{6,e/W}(p_1,p_2,p_3,p_4,p_5,p_6) = A_{4}^{(L)\mu}(p_2,p_3,p_4,\tilde{p}_W) \; L^{e/W}_{B,\mu}(p_1,p_5,p_6) \,, \label{eq:decayB}
\end{align}
where $p_W = p_5 + p_6$ and  $\tilde{p}_W = p_1 + p_5 + p_6$,
$L^{e}_{B,\mu}$ ($L^{W}_{B,\mu}$) is the decay current where the photon is emitted from the positron ($W$ boson), while $L_{A,\mu}$ is simply the $W^+ \to \nu e^+$ decay current.
The QCD corrections affect only the $W$-production amplitudes $A_{5,u/d}^{(L)\mu}$ and $A_{4}^{(L)\mu}$. We adopt the same decomposition for the finite remainders,
\begin{align}
& F^{(L)}_{6,u/d}(p_1,p_2,p_3,p_4,p_5,p_6) = F_{5,u/d}^{(L)\mu}(p_1,p_2,p_3,p_4,p_W) \; L_{A,\mu}(p_5,p_6) \,, \label{eq:decayAF}\\
& F^{(L)}_{6,e/W}(p_1,p_2,p_3,p_4,p_5,p_6) = F_{4}^{(L)\mu}(p_2,p_3,p_4,\tilde{p}_W) \; L^{e/W}_{B,\mu}(p_1,p_5,p_6) \,. \label{eq:decayBF}
\end{align}

In the next subsections we discuss the computation of $F_{5,d}^{(L)\mu}$ and $F_{4}^{(L)\mu}$. We recall that $F_{6,u}^{(L)}$ can be obtained from $F_{6,d}^{(L)}$ through the amplitude-level relation given in Eq.~\eqref{eq:ufromd}, which we rewrite here for the finite remainders,
\begin{equation}  \label{eq:ufromdF}
F_{6,u}^{(L)}(1^{h_1}_\gamma,2^+_{\bar{u}},3^{h_3}_g,4^-_d,5^{h_5}_{\nu},6^{h_6}_{\bar{e}})
=
F_{6,d}^{(L)}(1^{h_1}_\gamma,4^-_{\bar{u}},3^{h_3}_g,2^+_d,5^{h_5}_{\nu},6^{h_6}_{\bar{e}}) \,.
\end{equation}
We begin by describing how we parameterise the kinematics. Next, we discuss how we decompose the $W$-production five- and four-particle amplitudes, in Sections~\ref{sec:A5amplitude} and~\ref{sec:B4amplitude} respectively, using the projector method. Section~\ref{sec:Reconstruction} is devoted to the finite-field setup which we use to reconstruct the analytic expressions of the finite remainders as linear combinations of rational coefficients and monomials of independent special functions. In Section~\ref{sec:Simplification} we present a strategy which allows us to simplify dramatically the expressions of the rational coefficients. Finally, in Section~\ref{sec:Permutations} we discuss how our analytic results for the minimal set of independent finite remainders can be used efficiently to evaluate numerically all the contributions to the squared finite remainder summed over helicity and colour.

\subsection{Kinematics}
\label{sec:Kinematics}
In this section we describe the kinematics of the process $\wpaj$~\eqref{eq:wpaj}.
All the external momenta $p_i^{\mu}$ are massless,
\begin{equation}
p_i^2 = 0 \quad \forall \, i = 1,\ldots,6 \,,
\end{equation}
and taken to be outgoing, so that momentum conservation is
\begin{equation}
\sum_{i=1}^{6} p_i = 0 \,.
\end{equation}
We consider the external momenta $p_i^{\mu}$ to live in a four-dimensional Minkowski space. As a result there are eight independent scalar invariants, which we choose as
\begin{equation}
\vec{s}_{6} = \left\{ s_{12},s_{23},s_{34},s_{45},s_{56},s_{16},s_{123},s_{234}\right\} \,.
\label{eq:sijs6pt}
\end{equation}
It is also possible to form pseudo-scalar invariants by contracting the Levi-Civita symbol $\eps_{\mu \nu \rho \sigma}$ with any four external momenta. The six-particle kinematics is therefore fully determined by the scalar invariants $\vec{s}_6$ and by one pseudo-scalar invariant, which captures all the space-time parity information of the phase space. In analogy with the five-particle kinematics, we choose
\begin{align} \label{eq:tr5}
\tr_5 = 4 i \epsilon_{\mu \nu \rho \sigma} p_1^{\mu} p_2^{\nu} p_3^{\rho} p_4^{\sigma} \,.
\end{align}
The latter is related to the scalar invariants through
\begin{align} \label{eq:tr5squared}
\tr_5^2 =  \Delta_5 := \text{det}\left(2 p_i \cdot p_j \right)\bigl|_{i,j=1,\ldots,4} \,,
\end{align}
where the right-hand side is a degree-four polynomial in the scalar invariants. 

Only a subset of these invariants are relevant for the computation of $A_{5,d}^{(L)\mu}$, which has five-point kinematics with an external massive particle. We choose the following independent five-point scalar invariants for computing $A_{5,d}^{(L)\mu}$,
\begin{equation}
\vec{s}_{5} = \left\{ s_{12},s_{23},s_{34},s_{123},s_{234},s_{56}\right\} \,,
\label{eq:sijs5pt}
\end{equation}
together with $\tr_5$. Even fewer variables are relevant for $A_{4}^{(L)\mu}$. Since the latter has four-point kinematics with an external massive particle, no pseudo-scalar invariant can be formed and it is thus independent of $\tr_5$. Moreover, it depends only on three of the scalar invariants in $\vec{s}_5$. Nonetheless, we view it as a function of $\vec{s}_{5}$ in order to have a homogeneous setup.

When attaching the $W$-boson decay currents ($L_{A,\mu}$ and $L^{e/W}_{B,\mu}$) to $A_{5,d}^{(L)\mu}$ and $A_{4}^{(L)\mu}$ (see Eqs.~\eqref{eq:decayA}~and~\eqref{eq:decayB}) we find it convenient to describe the massless six-point kinematics using a parameterisation based on momentum twistors~\cite{Hodges:2009hk} (see e.g.\ Refs.~\cite{Badger:2013gxa,Badger:2016uuq} for a thorough discussion of momentum twistors in this context). We adopt the following momentum-twistor parameterisation,
\begin{align}
  Z =
  \begin{pmatrix}
    1 & 0 & y_1 & y_2 & y_3 & y_4\\
    0 & 1 & 1   & 1   & 1   & 1 \\
    0 & 0 & 0   & \tfrac{x_5}{x_2} & x_6 & 1 \\
    0 & 0 & 1   & 1   & x_7 & 1-\tfrac{x_8}{x_5}
  \end{pmatrix} \,,
  \label{eq:ZmatrixW}
\end{align}
where the columns give the four-component momentum twistors of the six external particles, and we used the short-hand notation $y_i = \sum_{j=1}^{i} \prod_{k=1}^{j} \frac{1}{x_k}$. The eight momentum-twistor variables $x_i$ are related to the external momenta through
\begin{align}
  x_1 &=  s_{12} \,, &
  x_2 &= -\frac{\trp(1234)}{s_{12}s_{34}} \,, &
  x_3 &= -\frac{\trp(1345)}{s_{45}s_{13}} \,, &
  x_4 &= -\frac{\trp(1456)}{s_{14}s_{56}} \,, \nonumber \\
  x_5 &=  \frac{s_{23}}{s_{12}} \,, &
  x_6 &= -\frac{\trp(15(3+4)2)}{s_{12}s_{15}} \,, &
  x_7 &=  \frac{\trp(51(2+3)(2+3+4))}{s_{15} s_{23}} \,, &
  x_8 &=  \frac{s_{123}}{s_{12}} \,, &
\label{eq:mtvardefs}
\end{align}
with $\tr_{\pm}(ij \cdots kl) = \tr((1\pm\gamma_5)\slashed{p}_i\slashed{p}_j \cdots\slashed{p}_k\slashed{p}_l)/2$. Note that the momentum-twistor variables $x_i$ in general transform in a non-trivial way under space-time parity. We implement the action of parity on momentum-twistor expressions as a change of momentum-twistor variables which leave unchanged the scalar invariants and flips the sign of $\tr_5$. The definition of the parity-flipped momentum-twistor variables can be obtained by trading $\trp$ for $\trm$ in Eqs.~\eqref{eq:mtvardefs}.

\subsection{Structure of the five-particle $W$-production amplitudes}
\label{sec:A5amplitude}
We decompose the five-point $W$-production amplitude $A_{5,d}^{(L)\mu}$ using the external momenta $(p_1,p_2,p_3,p_4)$ as the spanning basis,
\begin{align}
A_{5,d}^{(L)\mu} = p_1^{\mu} a^{(L)}_{1}  + p_2^{\mu} a^{(L)}_{2} + p_3^{\mu} a^{(L)}_{3} + p_4^{\mu} a^{(L)}_{4} \,.
\end{align}
The coefficients $a^{(L)}_{i}$ can be obtained by inverting the system of equations
\begin{align}
a^{(L)}_{i} = \sum_{j=1}^{4} \left(\Delta^{-1}\right)_{ij} \; \tilde{A}_{5,j}^{(L)} \,,
\label{eq:solveai}
\end{align}
where
\begin{align}
\label{eq:Delta}
&\Delta_{ij} = p_i\cdot p_j \,, \\
\label{eq:A5i}
&\tilde{A}_{5,i}^{(L)} = p_i \cdot A_{5,d}^{(L)} \,.
\end{align}
The \textit{contracted amplitudes} $\tilde{A}_{5,i}^{(L)}$ are computed by first generating the five-point process with an on-shell $W$ boson, followed by
replacing the $W$-boson polarisation vector by the four external momenta in the spanning basis, $(p_1,p_2,p_3,p_4)$.
We then apply tensor decomposition, taking into account the four-dimensional nature of the external states 
as proposed in Refs.~\cite{Peraro:2019cjj,Peraro:2020sfm}, to express each contracted amplitude $\tilde{A}_{5,i}^{(L)}$ as a linear combination of 8 independent tensor structures $\{T_j\}_{j=1}^8$,
\begin{align}
\tilde{A}_{5,i}^{(L)} = \sum_{j=1}^{8} T_{j} \alpha_{i,j}^{(L)} \,,
\end{align}
where
\begin{align}
T_{1}  & = \bar{u}(p_4) \slashed{p}_{1} v(p_2) \; p_{2}\cdot \vareps(p_1,q_1) \; p_{2}\cdot\vareps(p_3,q_3) \,, \nn
T_{2}  & = \bar{u}(p_4) \slashed{p}_{1} v(p_2) \; p_{2}\cdot \vareps(p_1,q_1) \; p_{4}\cdot\vareps(p_3,q_3) \,, \nn
T_{3}  & = \bar{u}(p_4) \slashed{p}_{1} v(p_2) \; p_{4}\cdot \vareps(p_1,q_1) \; p_{2}\cdot\vareps(p_3,q_3) \,, \nn
T_{4}  & = \bar{u}(p_4) \slashed{p}_{1} v(p_2) \; p_{4}\cdot \vareps(p_1,q_1) \; p_{4}\cdot\vareps(p_3,q_3) \,, \\
T_{5}  & = \bar{u}(p_4) \slashed{p}_{3} v(p_2) \; p_{2}\cdot \vareps(p_1,q_1) \; p_{2}\cdot\vareps(p_3,q_3) \,, \nn
T_{6}  & = \bar{u}(p_4) \slashed{p}_{3} v(p_2) \; p_{2}\cdot \vareps(p_1,q_1) \; p_{4}\cdot\vareps(p_3,q_3) \,, \nn
T_{7}  & = \bar{u}(p_4) \slashed{p}_{3} v(p_2) \; p_{4}\cdot \vareps(p_1,q_1) \; p_{2}\cdot\vareps(p_3,q_3) \,, \nn
T_{8}  & = \bar{u}(p_4) \slashed{p}_{3} v(p_2) \; p_{4}\cdot \vareps(p_1,q_1) \; p_{4}\cdot\vareps(p_3,q_3) \,. \nonumber
\end{align}
Here, $q_1$ and $q_3$ are arbitrary reference vectors for the photon and the gluon polarisation states, respectively.
We set $q_1=p_3$ and $q_3=p_1$ throughout our computation.
The tensor coefficients $\alpha^{(L)}_{i,j}$ can be obtained by
\begin{align}
\alpha^{(L)}_{i,j} = \sum_{k=1}^{8} \left(\Theta^{-1}\right)_{jk} \; \sum_{\text{pol}} {T}_{k}^\dagger \tilde{A}_{5,i}^{(L)} \,,
\end{align}
where
\begin{align}
\Theta_{ij}  = \sum_{\text{pol}} T_{i}^\dagger T_{j} \,,
\end{align}
and the polarisation-vector sum for the photon and gluon is
\begin{align}
\sum_{\text{pol}} \vareps^*_\mu(p_i,q_i) \vareps_\nu(p_i,q_i) = -g_{\mu\nu} + \frac{ p_{i\mu}q_{i\nu} + q_{i\mu}p_{i\nu} }{p_i \cdot q_i}, \qquad i=1,3\,.
\label{eq:polsum}
\end{align}
We further specify the helicity states of the spinors and polarisation vectors in the tensor structures $\{T_j\}_{j=1}^8$,
\begin{align}
T^{h_1 h_2 h_3 h_4}_{1} & = \bar{u}(p_4,h_4) \slashed{p}_{1} v(p_2,h_2) \; p_{2}\cdot \vareps(p_1,q_1,h_1) \; p_{2}\cdot\vareps(p_3,q_3,h_3) \,, \nn
T^{h_1 h_2 h_3 h_4}_{2} & = \bar{u}(p_4,h_4) \slashed{p}_{1} v(p_2,h_2) \; p_{2}\cdot \vareps(p_1,q_1,h_1) \; p_{4}\cdot\vareps(p_3,q_3,h_3) \,, \nn
T^{h_1 h_2 h_3 h_4}_{3} & = \bar{u}(p_4,h_4) \slashed{p}_{1} v(p_2,h_2) \; p_{4}\cdot \vareps(p_1,q_1,h_1) \; p_{2}\cdot\vareps(p_3,q_3,h_3) \,, \nn
T^{h_1 h_2 h_3 h_4}_{4} & = \bar{u}(p_4,h_4) \slashed{p}_{1} v(p_2,h_2) \; p_{4}\cdot \vareps(p_1,q_1,h_1) \; p_{4}\cdot\vareps(p_3,q_3,h_3) \,, \\
T^{h_1 h_2 h_3 h_4}_{5} & = \bar{u}(p_4,h_4) \slashed{p}_{3} v(p_2,h_2) \; p_{2}\cdot \vareps(p_1,q_1,h_1) \; p_{2}\cdot\vareps(p_3,q_3,h_3) \,, \nn
T^{h_1 h_2 h_3 h_4}_{6} & = \bar{u}(p_4,h_4) \slashed{p}_{3} v(p_2,h_2) \; p_{2}\cdot \vareps(p_1,q_1,h_1) \; p_{4}\cdot\vareps(p_3,q_3,h_3) \,, \nn
T^{h_1 h_2 h_3 h_4}_{7} & = \bar{u}(p_4,h_4) \slashed{p}_{3} v(p_2,h_2) \; p_{4}\cdot \vareps(p_1,q_1,h_1) \; p_{2}\cdot\vareps(p_3,q_3,h_3) \,, \nn
T^{h_1 h_2 h_3 h_4}_{8} & = \bar{u}(p_4,h_4) \slashed{p}_{3} v(p_2,h_2) \; p_{4}\cdot \vareps(p_1,q_1,h_1) \; p_{4}\cdot\vareps(p_3,q_3,h_3) \,, \nonumber
\end{align}
from which we obtain the contracted helicity amplitudes,
\begin{align}
\label{eq:A5Ldef}
\tilde{A}_{5,i}^{(L),h_1 h_2 h_3 h_4} = \sum_{j,k=1}^{8} T_{j}^{h_1 h_2 h_3 h_4} \left(\Theta^{-1}\right)_{jk} \tilde{\mathcal{A}}_{5,k \, i}^{(L)} \,,
\end{align}
with
\begin{align} \label{eq:Asumpol}
\tilde{\mathcal{A}}_{5,k \, i}^{(L)}  = \sum_{\text{pol}} {T}_{k}^\dagger \tilde{A}_{5,i}^{(L)} \,.
\end{align}

We carry out the same decomposition for the five-particle finite remainder $F_{5,d}^{(L)\mu}$, arriving at the following formula for the contracted helicity finite remainders,
\begin{align}
\label{eq:FA5Ldef}
\tilde{F}_{5,i}^{(L),h_1 h_2 h_3 h_4} = \sum_{j,k=1}^{8} T_{j}^{h_1 h_2 h_3 h_4} \left(\Theta^{-1}\right)_{jk}  \, \tilde{\mathcal{F}}_{5,k \, i}^{(L)}\,,
\end{align}
where
\begin{align} \label{eq:FA5Ldefpt2}
\tilde{\mathcal{F}}_{5,k \, i}^{(L)}  = \sum_{\text{pol}} {T}_{k}^\dagger  \, p_{i \, \mu} \, F_{5,d}^{(L) \mu}  \,.
\end{align}

As discussed in Section~\ref{sec:amp}, the independent helicity configurations which we need to compute are
\begin{equation} \label{eq:indephel}
\left\{
\tilde{F}_{5,i}^{(L),+++-}, 
\tilde{F}_{5,i}^{(L),-++-}, 
\tilde{F}_{5,i}^{(L),+-++}, 
\tilde{F}_{5,i}^{(L),--++}
\right\} \,.
\end{equation}
We note that it is possible to compute directly the contracted finite remainders $\tilde{F}^{(L)}_{5,i}$ without specifying the helicity states. 
In our setup, however, such a computation would lead to more complicated analytic expressions compared to the results obtained for the contracted helicity amplitudes.

\subsection{Structure of the four-particle $W$-production amplitudes}
\label{sec:B4amplitude}

The four-particle $W$-production amplitude $A_{4}^{(L)\mu}$ has been computed in the context of $W+1j$ production at the LHC ($q\bar{q} \to Wg$)~\cite{Gehrmann:2011ab}, which is
a crossing of the $e^+ e^- \to q\bar{q}g$ amplitude~\cite{Garland:2002ak}. In our case it is convenient to express $A_{4}^{(L)\mu}$ in terms of the same special function basis as $A_{5,d}^{(L)\mu}$. This guarantees a uniform combination of the different contributions to the full amplitude. We therefore re-derive the $A_{4}^{(L)\mu}$ amplitude using our computational framework.
We decompose the $A_{4}^{(L)\mu}$ amplitude using the following tensor structures~\cite{Garland:2002ak},
\begin{equation}
A_{4}^{(L)\mu}(p_2,p_3,p_4) = \sum_{i=1}^{7} b_i^{(L)} Y_{i}^\mu \,,
\label{eq:B4decomposition}
\end{equation}
where
\begin{align}
  Y_{1}^{\mu}  &= \bar{u}(p_4) \slashed{p}_3 v(p_2) \, \vareps_3 \cdot p_4 \, p_4^\mu 
                    -p_3 \cdot p_4 \, \bar{u}(p_4) \slashed{\vareps}_3 v(p_2) \, p_4^\mu \nn
                 & \quad -(p_2 \cdot p_4 + p_3 \cdot p_4) \left[\bar{u}(p_4) \slashed{p}_3 v(p_2) \, \vareps_3^\mu - \bar{u}(p_4) \slashed{\vareps}_3 v(p_2) \, p_3^\mu \right] \,, \nn
  Y_{2}^{\mu}  &= \bar{u}(p_4) \slashed{p}_3 v(p_2) \, \vareps_3 \cdot p_4 \, p_3^\mu 
                    -p_3 \cdot p_4 \, \bar{u}(p_4) \slashed{\vareps}_3 v(p_2) \, p_3^\mu \nn
                 & \quad -(p_2 \cdot p_3 + p_3 \cdot p_4) \left[\bar{u}(p_4) \slashed{p}_3 v(p_2) \, \vareps_3^\mu - \bar{u}(p_4) \slashed{\vareps}_3 v(p_2) \, p_3^\mu \right] \,, \nn
  Y_{3}^{\mu}  &= \bar{u}(p_4) \slashed{p}_3 v(p_2) \, \vareps_3 \cdot p_4 \, p_2^\mu 
                    -p_3 \cdot p_4 \, \bar{u}(p_4) \slashed{\vareps}_3 v(p_2) \, p_2^\mu \nn
                 & \quad -(p_2 \cdot p_3 + p_2 \cdot p_4) \left[\bar{u}(p_4) \slashed{p}_3 v(p_2) \, \vareps_3^\mu - \bar{u}(p_4) \slashed{\vareps}_3 v(p_2) \, p_3^\mu \right] \,, \label{eq:4ptDecomposition}\\
  Y_{4}^{\mu}  &= \bar{u}(p_4) \slashed{p}_3 v(p_2) \, \vareps_3 \cdot p_2 \, p_4^\mu 
                    -p_2 \cdot p_3 \, \bar{u}(p_4) \slashed{\vareps}_3 v(p_2) \, p_4^\mu \,, \nn
  Y_{5}^{\mu}  &= \bar{u}(p_4) \slashed{p}_3 v(p_2) \, \vareps_3 \cdot p_2 \, p_3^\mu 
                    -p_2 \cdot p_3 \, \bar{u}(p_4) \slashed{\vareps}_3 v(p_2) \, p_3^\mu \,, \nn
  Y_{6}^{\mu}  &= \bar{u}(p_4) \slashed{p}_3 v(p_2) \, \vareps_3 \cdot p_2 \, p_2^\mu 
                    -p_2 \cdot p_3 \, \bar{u}(p_4) \slashed{\vareps}_3 v(p_2) \, p_2^\mu \,, \nn
  Y_{7}^{\mu}  &=   p_3 \cdot p_4 \, \bar{u}(p_4) \gamma^\mu v(p_2) \, \vareps_3 \cdot p_2
                    - p_2 \cdot p_3 \, \bar{u}(p_4) \gamma^\mu v(p_2) \, \vareps_3 \cdot p_4 \nn
                 & \quad -(p_2 \cdot p_3) \left[ \bar{u}(p_4) \slashed{p}_3 v(p_2) \, \vareps_3^\mu  - \bar{u}(p_4) \slashed{\vareps}_3 v(p_2) \, p_3^\mu \right] \,. \nonumber
\end{align}
The coefficients $b_i^{(L)}$ can be determined through
\begin{equation}
b^{(L)}_{i} = \sum_{j} \left(\Omega^{-1}\right)_{ij}  \tilde{A}^{(L)}_{4,j} \,,
\label{eq:solvebi}
\end{equation}
where
\begin{align}
\label{eq:Omega}
& \Omega_{ij} = \sum_{\text{pol}} Y^{\mu\dagger}_{i} Y_{j\mu} \,, \\
\label{eq:B4Ldef}
& \tilde{A}^{(L)}_{4,i} = \sum_{\text{pol}} Y^{\mu\dagger}_{i} A^{(L)}_{4 \, \mu} \,.
\end{align}
The gluon polarisation-vector sum follows from Eq.~\eqref{eq:polsum}.
We note that the tensor structures in Eq.~\eqref{eq:4ptDecomposition} are different from the ones employed in Ref.~\cite{Garland:2002ak}.
Here, we start from 12 tensor structures that are linearly independent in 4 dimensions~\cite{Peraro:2019cjj,Peraro:2020sfm} and reduce them to 7 by imposing Ward identities.
Since $A^{(L)\mu}_4$ is a four-point amplitude, it does not depend on $\tr_5$. For the sake of uniformity, we express it in terms of the five-point Mandelstam invariants $\vec{s}_5$~\eqref{eq:sijs5pt}. 
In contrast to the computation of $A^{(L)\mu}_{5,d}$, here we derive the contracted amplitudes $\tilde{A}^{(L)}_{4,i}$ directly without specifying the helicity states, since the four-point computation
is relatively simple. The helicity states for the tensor structures $Y_\mu^{i}$ are specified when the decay currents are attached, following Eq.~\eqref{eq:decayB}.

Once again we perform the same decomposition on the corresponding four-particle finite remainder $F^{(L)\mu}_{4}$. The resulting formula for the contracted finite remainders is
\begin{align} \label{eq:FB4Ldef}
\tilde{F}^{(L)}_{4,i} = \sum_{\text{pol}} Y^{\mu\dagger}_{i} F^{(L)}_{4 \, \mu} \,.
\end{align}

\subsection{Amplitude reduction and analytic reconstruction}
\label{sec:Reconstruction}

In this section we present the analytic computation of the contracted five- and four-particle finite remainders, $\tilde{F}_{5,i}^{(L),h_1 h_2 h_3 h_4}$~\eqref{eq:FA5Ldef} and $\tilde{F}_{4,i}^{(L)}$~\eqref{eq:FB4Ldef} respectively, at one and two loops.
We adopt the framework used in Refs.~\cite{Badger:2021nhg,Badger:2021owl,Badger:2021imn,Badger:2021ega}, based on Feynman diagrams and functional reconstruction from numerical sampling over finite field. For the latter we employ the \textsc{FiniteFlow} framework~\cite{Peraro:2019svx}.

In order to use the finite field technique, we need to have a rational parameterisation of the
kinematics. As discussed in Section~\ref{sec:Kinematics}, the five-particle phase space is described
by the six scalar invariants $\vec{s}_5$~\eqref{eq:sijs5pt} together with the pseudo-scalar
invariant $\tr_5$~\eqref{eq:tr5}. It follows from Eq.~\eqref{eq:tr5squared} that $\tr_5$ is given by
the square root of $\Delta_5$, up to the overall sign which encodes the parity information. This
square root may be dealt with in two ways: either parameterise the kinematics to explicitly
rationalise $\sqrt{\Delta}_5$, or parameterise the finite remainders
$\tilde{F}_{5,i}^{(L),h_1h_2h_3h_4}$ such that the dependence on $\tr_5$ is analytic. We choose the
latter option in this case. The pseudo-scalar invariant $\tr_5$ can enter the computation in three
distinct ways. First, it can originate from the $\gamma_5$ in the axial coupling of the $W$ boson.
As we discussed in Section~\ref{sec:amp}, we set up the computation so that only the vector coupling
of $W$ is used. Second, $\tr_5$ is in general needed to capture the parity-odd part of the
spinor-helicity expressions. However, the pseudo-scalar invariant enters the contracted helicity
finite remainder in Eq.~\eqref{eq:FA5Ldef} only through the spanning basis elements $T_{j}^{h_1 h_2
h_3 h_4}$, which are known analytically and do not need to be reconstructed over finite fields.
Finally, $\tr_5$ is present in the definition of the canonical master integrals, which also
introduces three other square roots, $\sqrt{\Delta_3^{(i)}}$ with $i=1,2,3$, relevant for the
analytic structure of the Feynman integrals~\cite{Abreu:2020jxa,Chicherin:2021dyp}. Since the square
roots appear only as overall normalisation factors, we can re-absorb them in the definition of the
canonical master integrals, and thus reduce the finite remainders to master integrals which are all
manifestly scalar. With this setup the only parts of the contracted finite remainders which need to
be reconstructed are rational functions of the scalar invariants $\vec{s}_5$ only, and can thus be
sampled numerically over finite fields.

The starting point of our computation is the expression of the $W$-production five- and
four-particle amplitudes, $A_{5,d}^{(L)\mu}$ and $A_4^{(L)\mu}$, in terms of Feynman diagrams, which
we generate using \textsc{Qgraf}~\cite{Nogueira:1991ex}. For those interested in counting Feynman
diagrams, there are 20 diagrams for $A_{5,d}^{(1),1 \mu}$, 231 diagrams for $A_{5,d}^{(2),1 \mu}$, 32 diagrams for $A_{5,d}^{(2),n_f \mu}$, 
7 diagrams for $A_{4}^{(1),1 \mu}$, 74 diagrams for $A_{4}^{(2),1 \mu}$, and 13 diagrams for
$A_{4}^{(2),n_f \mu}$. Clearly this signifies nothing more than that the number of diagrams is not a
good measure of complexity. We want to obtain analytic, IBP-compatible expressions for the contracted amplitudes, $\tilde{A}_{5,i}^{(L),h_1 h_2 h_3 h_4}$ and $\tilde{A}_{4,i}^{(L)}$. For the four-point amplitude $A^{(L)\mu}_{4}$ we apply the projectors and sum over all polarisation states as in Eq.~\eqref{eq:B4Ldef}. For the five-point amplitude $A^{(L)\mu}_{5,d}$ we contract by the external momenta and apply the projectors, as in Eqs.~\eqref{eq:A5i} and~\eqref{eq:Asumpol}, respectively. We then rewrite the resulting expressions in terms of scalar Feynman integrals belonging to the master topologies defined in Refs.~\cite{Badger:2021nhg,Badger:2021ega}.
We carry out all these operations analytically using \textsc{Mathematica} and \textsc{Form}~\cite{Kuipers:2012rf,Ruijl:2017dtg} scripts. As a result, we obtain analytic expressions for $\tilde{\mathcal{A}}_{5,i\, k}^{(L)}$~\eqref{eq:Asumpol} and $\tilde{A}_{4,i}^{(L)}$~\eqref{eq:B4Ldef} as linear combinations of scalar Feynman integrals with rational coefficients functions of $\vec{s}_5$. 
In order to obtain the contracted helicity-amplitudes $\tilde{A}_{5,i}^{(L), h_1h_2h_3h_4}$ from the $\tilde{\mathcal{A}}_{5,i\, k}^{(L)}$'s we further need to multiply by the spanning basis elements $T^{h_1h_2h_3h_4}_{j}$ and by the inverse of $\Theta$, as shown in Eq.~\eqref{eq:A5Ldef}. We do these operations (including the inversion of $\Theta$) numerically within the finite field framework.

We reduce the scalar integrals to the canonical master integrals identified in Ref.~\cite{Abreu:2020jxa}, which we modified so as to re-absorb the square roots. We generate the IBP relations~\cite{Chetyrkin:1981qh} using \textsc{LiteRed}~\cite{Lee:2012cn} in \textsc{Mathematica}, and solve them numerically over finite fields using the Laporta algorithm~\cite{Laporta:2000dsw} through \textsc{FiniteFlow}'s linear solver. We then perform a Laurent expansion of the rational coefficients around $\eps=0$, and map the canonical master integrals onto square roots and the special function basis of  Ref.~\cite{Badger:2021nhg} up to the required order in $\eps$. We label the special function basis by $\{f_k\}$. We truncate the $\eps$ expansion at order $\eps^2$ at one loop and at order $\eps^0$ at two loops. Finally, we subtract the UV/IR poles as in Eq.~\eqref{eq:finiteremainder} and define the contracted finite remainders, which we represent as
\begin{align} \label{eq:finrem_ratcoeffs}
\begin{aligned}
& \tilde{F}_{5,i}^{(L), h_1h_2h_3h_4} = \Phi^{h_1h_2h_3h_4}_{5} \sum_j \left[ q_{i,j}^{h_1h_2h_3h_4}\left(\vec{s}_5\right) + \tr_5 \, r_{i,j}^{h_1h_2h_3h_4} \left(\vec{s}_5\right) \right] \text{mon}_j\left(\tr_5,\sqrt{\Delta_3^{(l)}}, \{f_k\} \right) \,, \\
& \tilde{F}_{4,i}^{(L)} = \sum_j t_{i,j}\left(\vec{s}_5\right) \text{mon}_j\left(\tr_5,\sqrt{\Delta_3^{(l)}}, \{f_k\} \right) \,, \\
\end{aligned}
\end{align}
where $\text{mon}_j(x,y,\ldots)$ denotes monomials in $x,y,\ldots$, while $q_{i,j}^{h_1h_2h_3h_4}$, $r_{i,j}^{h_1h_2h_3h_4}$ and $t_{i,j}$ are rational functions of $\vec{s}_5$. Note that we pull out from the five-particle finite remainders an arbitrary phase factor $\Phi^{h_1h_2h_3h_4}_{5}$ carrying all the helicity weights, so that the coefficients $q_{i,j}^{h_1h_2h_3h_4}$ and $r_{i,j}^{h_1h_2h_3h_4}$ are scalar and hence functions of $\vec{s}_5$ only. We recall that the helicity configuration is assigned to the four-particle finite remainders when attaching the decay current, as discussed in Section~\ref{sec:B4amplitude}. The cancellation of the poles at this stage provides a robust check of the result prior to the rational reconstruction. Furthermore, it typically leads to simplifications which make the finite remainders easier to reconstruct than the bare amplitudes. This chain of operations is implemented in the \textsc{FiniteFlow} framework, and ultimately amounts to an algorithm which samples numerically over finite fields the rational coefficients in the contracted finite remainders.

Finally, we need to reconstruct the rational coefficients of the contracted finite remainders from their numerical values. Following Refs.~\cite{Abreu:2018zmy,Badger:2021nhg,Badger:2021imn,Badger:2021ega}, we perform a number of optimisations to reduce the number of required sample points. We follow the strategy outlined in Ref.~\cite{Badger:2021imn}. First of all, we set $s_{12}=1$. We recover the analytic dependence on $s_{12}$ a posteriori through dimensional analysis. Second, we fit the $\mathbb{Q}$-linear relations among the rational coefficients, and solve them so as to express the most complicated coefficients in terms of the simplest ones. Third, we reconstruct the coefficients on a random univariate phase-space slice modulo a large prime number, and match them with ans\"atze made of the following factors (with $s_{12}=1$):
\begin{align} \label{eq:coeffansatz}
\begin{aligned} 
\biggl\{ & s_{12}\,, s_{23}\,, s_{34}\,, s_{23} + s_{34}\,, s_{23} - s_{234}\,, s_{234} - s_{34}\,, s_{123} - s_{56}\,, s_{234} - s_{56}\,, s_{12} - s_{123} + s_{23}\,, \\
& s_{12} + s_{234} - s_{34} \,, s_{23} - s_{234} + s_{34}\,, s_{12} + s_{234} - s_{56}\,, s_{12} - s_{123} - s_{34}\,, s_{123} s_{234} - s_{23} s_{56}\,, \\
& s_{12} - s_{123} + s_{23} - s_{34} \,, s_{123} - s_{23} + s_{234} - s_{56} \,, s_{12} + s_{234} - s_{34} - s_{56} \,, \\
& s_{12} s_{234} - s_{123} s_{234} + s_{23} s_{234} - s_{234} s_{34} + s_{34} s_{56} \,, \\
& s_{12} s_{234} - s_{123} s_{234} - s_{234} s_{34} + s_{23} s_{56} + s_{34} s_{56} \,, s_{12} s_{234} + s_{234}^2 - s_{234} s_{34} - s_{234} s_{56} + s_{34} s_{56} \,, \\
& s_{12} s_{123} + s_{123} s_{234} - s_{123} s_{34} - s_{12} s_{56} - s_{23} s_{56} \,, s_{12} s_{123} - s_{123}^2 - s_{123} s_{34} - s_{12} s_{56} + s_{123} s_{56} \,, \\
& s_{12}^2 - s_{12} s_{123} + s_{12} s_{234} - s_{123} s_{234} - 2 s_{12} s_{34} + s_{123} s_{34} - s_{234} s_{34} + s_{34}^2 + s_{23} s_{56} \,, \\
& s_{12} s_{23} - s_{12} s_{234} + s_{23} s_{234} - s_{234}^2 - s_{23} s_{34} + s_{234} s_{34} - s_{23} s_{56} + s_{234} s_{56} - s_{34} s_{56} \,, \\
& \lambda\left(s_{12}, s_{34}, s_{56} \right) \,, \lambda\left(s_{23}, s_{14}, s_{56}\right) \,, \tr_5^2 \biggr\} \,,
\end{aligned}
\end{align}
where $\lambda$ is the K\"allen function,
\begin{align}
\lambda(a,b,c) = a^2 + b^2 + c^2 - 2 a b - 2 b c - 2 c a \,.
\end{align}
The factors in the list~\eqref{eq:coeffansatz} are a subset of the letters of the (planar) one-mass pentagon alphabet~\cite{Abreu:2020jxa}, namely of the arguments of the logarithmic integration kernels appearing in the differential equations satisfied by the master integrals. The letters govern the singularity structure of the master integrals and hence of the special function basis $\{f_k\}$. It is therefore natural to expect that the denominators of the rational coefficients multiplying the special functions should factorise in terms of letters, and indeed the previous experience has shown that this is the case~\cite{Abreu:2018zmy,Badger:2021nhg,Badger:2021imn,Badger:2021ega}. It follows that we can determine entirely the denominators of the rational coefficients by matching them against ans\"atze made of the factors in Eq.~\eqref{eq:coeffansatz} on a univariate slice. Part of the numerators may in general be caught by this approach as well. In Table~\ref{tab:degrees} we show the impact of this strategy on the highest polynomial degrees of the rational coefficients which need to be reconstructed for the five-particle contracted finite remainders. Note that we process all helicity configurations of the five-particle finite remainders simultaneously, but for the $n_f^0$ ones we separate the contractions by the external momenta into two subsets, $\{p_1,p_2\}$ and $\{p_3,p_4\}$, to reduce the memory usage. After this optimisation is done, the rational coefficients are reconstructed using the multivariate functional reconstruction algorithms implemented in \textsc{FiniteFlow}~\cite{Peraro:2019svx}.
\begin{table}[t!]
\begin{center}
\begin{tabular}{l|c|c|c}
 & $s_{12}=1$ & linear relations & factor matching \\
\hline
$\tilde{F}_{5,i}^{(2),1\, h_1 h_2 h_3 h_4}$ with $i=1,2$     & $44/44$ & $41/40$ & $41/0$ \\
$\tilde{F}_{5,i}^{(2),1\, h_1 h_2 h_3 h_4}$ with $i=3,4$     & $48/47$ & $42/42$ & $42/0$ \\
$\tilde{F}_{5,i}^{(2),n_f\, h_1 h_2 h_3 h_4}$ with $i=1,2,3,4$ & $39/38$ & $26/24$ & $26/0$ \\
\end{tabular}
\end{center}
\caption{Maximal total polynomial degrees of the rational coefficients of the contracted two-loop five-particle finite remainders at each stage of the optimisation procedure for the finite-field reconstruction, in the form numerator/denominator. The coefficients are  functions of the five scalar invariants $\{s_{23},s_{34},s_{123},s_{234},s_{56}\}$ ($s_{12}=1$). The independent helicity configurations~\eqref{eq:indephel} are processed simultaneously, while the contractions by the external momenta for $\tilde{F}_{5,i}^{(2),1\, h_1 h_2 h_3 h_4}$ are separated into two subsets to reduce the memory usage.}
\label{tab:degrees}
\end{table}

\subsection{Simplification of the rational coefficients}
\label{sec:Simplification}

The resulting analytic expressions of the rational coefficients of the finite remainders are rather bulky. The standard approach to simplify them relies on partial fraction decomposition, either multivariate~\cite{Leinartas:1978,Raichev:2012,Abreu:2019odu,Boehm:2020ijp,Heller:2021qkz,Bendle:2021ueg} or univariate with respect to a suitable variable~\cite{Badger:2021nhg,Badger:2021imn,Badger:2021ega,Abreu:2021asb}. For the rational coefficients of the four-point finite remainders, $t_{i,j}\left(\vec{s}_5\right)$ in Eq.~\eqref{eq:finrem_ratcoeffs}, we achieve a satisfactory simplification by performing a multivariate partial fraction decomposition with the \textsc{Mathematica} package \textsc{MultivariateApart}~\cite{Heller:2021qkz}, enhanced by \textsc{Singular}~\cite{DGPS} for the computation of the Gr\"obner bases. 

The rational coefficients of the five-particle finite remainders are instead substantially more involved. In order to simplify them, we look for a parameterisation of the five-particle kinematics leading to more compact expressions than the scalar invariants $\vec{s}_5$~\eqref{eq:sijs5pt}. We investigate how the complexity of the expressions varies when using momentum-twistor parameterisations~\cite{Hodges:2009hk}. The pseudo-scalar invariant $\tr_5$ is given by a rational function in terms of momentum-twistor variables, and we can thus add up the two terms of the coefficients of the special function monomials,
\begin{align}
\left( q_{i,j}^{h_1h_2h_3h_4}\left(\vec{s}_5\right) + \tr_5 \, r_{i,j}^{h_1h_2h_3h_4}\left(\vec{s}_5\right) \right) \biggl|_{\vec{s}_5 = \vec{s}_5(\vec{z})} = u_{i,j}^{h_1h_2h_3h_4}\left(\vec{z}\right) \,,
\end{align}
where by $\vec{z} = \{z_i \}_{i=1,\ldots,6}$ we denote generally the independent momentum-twistor variables. In particular, we consider the parameterisation proposed in Ref.~\cite{Badger:2021ega},
\begin{equation}
\begin{alignedat}{2}
  & z_1 = s_{12} \,, \qquad && z_2 = -\frac{\trp(1234)}{s_{12}s_{34}} \,, \\
  & z_3 = \frac{\trp(1341(5+6)2)}{s_{13}\trp{(14(5+6)2)}} \,, \qquad && z_4 = \frac{s_{23}}{s_{12}} \,, \\
  & z_5 = -\frac{\trm(1(2+3)(1+5+6)(5+6)23)}{s_{23}\trm(1(5+6)23)} \,, \qquad && z_6 = \frac{s_{456}}{s_{12}} \,.
\end{alignedat}
\label{eq:mtvardefs5pt}
\end{equation}
In previous applications, such a parameterisation has been used globally, i.e.\ in all amplitudes/finite remainders
irrespective of their helicity configuration. We find that this approach does not perform well in
this case, and does not lead to a major simplification in comparison with the expressions in terms of scalar invariants $\vec{s}_5$ and $\tr_5$. 
The rational parameterisation has the effect of breaking some symmetries in the kinematic quantities,
which results in some configurations being simpler than others. There is no reason for the parameterisation
to be a global choice, and in this case we exploit this fact and consider different parameterisations for each helicity configuration.

In practice, we consider all parameterisations which are obtained by permuting the massless momenta on the right-hand side of Eqs.~\eqref{eq:mtvardefs5pt}. For each helicity configuration we determine which permutations of the parameterisation lead to the most compact expression of the finite remainder at one loop. We then use them at two loops, and select the one which results in the simplest expression. We perform the change of variables over finite fields within the \textsc{FiniteFlow} framework, and measure the ``simplicity'' of the rational coefficients in terms of their numerator/denominator polynomial degrees, which can be determined without reconstructing the expression of the coefficients in terms of the new variables. Once the ``best'' parameterisation $\vec{s}_5 = \vec{s}_5 (\vec{z})$ is chosen for each helicity configuration, we reconstruct the analytic expression of the coefficients in terms of the new variables $\vec{z}$. For this purpose we make use of the finite field algorithm for univariate partial fraction decomposition presented in Refs.~\cite{Badger:2021nhg,Badger:2021imn}. We choose the variable to partial fraction with respect to so as to minimise the polynomial degrees of the separate terms of the decomposition. Breaking down the coefficients into univariate partial fractions simplifies the subsequent multivariate partial fraction decomposition, which we perform using \textsc{MultivariateApart}~\cite{Heller:2021qkz} enhanced with \textsc{Singular}~\cite{DGPS}. We apply it to each term of the univariate partial fraction decomposition separately, which is convenient as each term is by itself much simpler than the full coefficient. This is possible because \textsc{MultivariateApart}'s algorithm commutes with summation by design. The spurious poles introduced by the univariate partial fraction decomposition therefore cancel out after the multivariate partial fraction decomposition. In summary, our algorithm for the simplification of the rational coefficients of the five-particle finite remainders is the following.
\begin{enumerate}
\item Try all permutations of a given momentum-twistor parameterisation on the one-loop expressions and select the ones which lead to the lowest polynomial degrees.
\item Apply the parameterisations selected at step 1 on the two-loop rational coefficients and choose the one which leads to the lowest polynomial degree.
\item Decompose the two-loop rational coefficients in terms of the new variables into univariate partial fractions with respect to the variable which leads to the lowest polynomial degrees in the separate terms.
\item Decompose into multivariate partial fractions the separate terms of the univariate partial fractions using the algorithm of Ref.~\cite{Heller:2021qkz}, and sum them up cancelling the spurious poles.
\end{enumerate}
In hindsight, the first three steps could have been implemented within the original \textsc{FiniteFlow} setup. We did not attempt this approach because we did not need any further optimisation to reconstruct the rational coefficients of the $W$-production five-particle finite remainders. However, we believe that this strategy may also help to improve the rational reconstruction.

We apply this procedure separately on each of the helicity configurations, leading to different parameterisations for each of them.
The resulting expressions for the coefficients are remarkably more compact than the original ones in terms of the scalar invariants $\vec{s}_5$. For the most complicated finite remainder we achieved a compression in the file size of more than two orders of magnitude. The evaluation time of the rational coefficients is similarly improved.

\subsection{Numerical evaluation and permutations of the amplitudes}
\label{sec:Permutations}

In order to obtain the values of all the amplitudes in all the possible scattering channels we need to evaluate the minimal set of independent objects we reconstructed for different permutations of the external momenta (see e.g.\ Eq.~\eqref{eq:ufromd} for an explicit example). In this subsection we discuss how we implement this operation in an efficient way at the level of the numerical evaluation.

We denote a generic permutation of the external momenta by
\begin{align}
\sigma = \left(\sigma_1 \sigma_2 \sigma_3 \sigma_4\sigma_5 \sigma_6 \right) \,,
\end{align}
where the $\sigma_i$'s take distinct values in $\{1,2,3,4,5,6\}$, such that the action of $\sigma$ on an external momentum is given by
\begin{align}
\sigma \circ p_i = p_{\sigma_i} \,.
\end{align}
Not all $S_6$ permutations of $\{1,2,3,4,5,6\}$ are needed for this application. The required permutations belong to the subset $S_4 \times Z_2$, i.e.\ 
they are obtained by composing an $S_4$ permutation of $\{p_1,p_2,p_3,p_4\}$ and a $Z_2$ exchange of  $\{p_5,p_6\}$. In particular, $p_5$ and $p_6$ need to be exchanged in order to obtain the $\wmaj$ amplitudes according to Eq.~\eqref{eq:wmajamps}. Only the $S_4$ permutations are relevant for the $W$-production amplitudes (and hence for the special functions), since $p_5$ and $p_6$ enter them only in the sum $p_5+p_6$ (see e.g.\ Eqs.~\eqref{eq:decayA} and~\eqref{eq:decayB}). The $Z_2$ exchange is relevant only for the leptonic currents ($L_{A,\mu}$ and $L^{e/W}_{B,\mu}$), which are rational functions.

Given a generic amplitude/finite remainder $A$, function of the external momenta $\{p_i\}$, we define its permutation $\sigma$ as
\begin{align}
\left(\sigma \circ A\right)\left(\{p_i\} \right) = A \left( \{ \sigma \circ p_i \}  \right) \,.
\end{align}
In other words, we can obtain the value of the permuted amplitude by evaluating the amplitude in the original orientation of the external momenta at a permuted phase space point. While this operation is trivial for the rational functions, it is in general very subtle for the special functions. The reason is that a permutation in general maps the phase-space point to a different scattering region. This would require a complicated analytic continuation, since the special functions have a very intricate branch cut structure.

One way to overcome this problem is to evaluate the special functions numerically using the generalised series expansion method~\cite{Moriello:2019yhu}, implemented in the public \textsc{Mathematica} package \textsc{DiffExp}~\cite{Hidding:2020ytt}, as done in Refs.~\cite{Badger:2021nhg,Badger:2021ega}. Within this method the analytic continuation can be carried out systematically. This approach however requires that, for each phase-space point where we want to evaluate the permuted amplitudes, we evaluate the special functions at as many points as the number of needed permutations. 

For phase-space points in the physical scattering region we can adopt a much more efficient evaluation strategy: we use the \texttt{C++} package \textsc{PentagonFunctions++}~\cite{Chicherin:2021dyp}, which allows us to evaluate in the physical scattering region a larger basis of special functions, named one-mass pentagon functions. We denote them by $\{g_i\}$.
For this purpose we translate the special function basis $\{f_i\}$ of Ref.~\cite{Badger:2021nhg} to the one-mass pentagon function basis $\{g_i\}$ implemented in \textsc{PentagonFunctions++}. The translation takes the form
\begin{align} \label{eq:translation}
f_i = \sum_{j} w_{ij} \, \text{mon}_j \left(\{g_k\} \right) \,,
\end{align}
where $w_{ij} \in \mathbb{Q}$, and the sum runs over all the required monomials of the one-mass pentagon functions $\{g_k\}$. We obtain the transformation rules~\eqref{eq:translation} by matching the expressions of the master integrals in terms of special special functions given in Ref.~\cite{Badger:2021nhg} with that of Ref.~\cite{Chicherin:2021dyp}. The advantage of the one-mass pentagon functions with respect to the function basis $\{f_i\}$ or Ref.~\cite{Badger:2021nhg} is that their evaluation through the package \textsc{PentagonFunctions++} is extremely efficient, and their design allows us to generate the values of all $S_4$ permutations of the functions from those at the un-permuted phase-space point. 
The one-mass pentagon function basis $\{g_i\}$ is in fact closed under $S_4$ permutations. This means that, for any $\sigma \in S_4$, we can express the permuted one-mass pentagon functions evaluated at a given phase-space point as a combination of un-permuted pentagon functions evaluated at the same point,
\begin{align} \label{eq:permf}
\left(\sigma \circ g_i\right)(\vec{s}_5, \tr_5) = \sum_j \Sigma_{ij}^{(\sigma)} \text{mon}_j \left[ \{ g_k\left(\vec{s}_5, \tr_5 \right) \} \right] \,,
\end{align}
where $\Sigma_{ij}^{(\sigma)} \in \mathbb{Q}$, and we spelled out the dependence on the kinematics for the sake of clarity. These transformation rules are provided in Ref.~\cite{Chicherin:2021dyp}.\footnote{Note that Ref.~\cite{Chicherin:2021dyp} has a different labelling of the external momenta. Moreover, the package \textsc{PentagonFunctions++} works in a specific physical scattering region (the $s_{45}$ channel using the notation of Ref.~\cite{Chicherin:2021dyp}). A relabelling and a further permutation of the momenta are required to use \textsc{PentagonFunctions++} in the scattering region relevant for our application. We implemented these operations in the \textsc{Mathematica} evaluation script provided in the ancillary files, and refer to the original work~\cite{Chicherin:2021dyp} for a discussion of how to use \textsc{PentagonFunctions++} in a physical region different from the default one.} This strategy is advantageous because it minimises the number of evaluations of the special functions, which is the most time-consuming step in the  numerical evaluation of the colour and helicity summed squared amplitudes.

It is worth highlighting the special behaviour of the pseudo-scalar invariant $\tr_5$ in this chain of operations. In the physical scattering regions the reality of the momenta implies that $\tr_5^2 < 0$. In other words, $\tr_5$ is purely imaginary. The library \textsc{PentagonFunctions++} always assumes that $\text{Im}\left[\tr_5\right]>0$. The sign of $\tr_5$ however may change upon the action of an odd-signature permutation,
\begin{align}
\sigma \circ \text{tr}_5 = \text{sign}(\sigma) \, \text{tr}_5 \,,
\end{align}
or space-time parity.
The values of the one-mass pentagon functions for a negative imaginary part of $\tr_5$ can be obtained by flipping the sign of a subset of functions specified in Ref.~\cite{Chicherin:2021dyp}. In our setup however we do not need to. As discussed in Section~\ref{sec:Reconstruction}, we reduce the amplitudes to manifestly scalar master integrals, and group together the special functions and the square roots arising from the definition of the canonical master integrals. As a result, the monomials of special functions and square roots in the finite remainders~\eqref{eq:finrem_ratcoeffs} are scalar as well. Any sign change in the pentagon functions due to permutations or space-time parity is therefore compensated by that of the accompanying factor of $\tr_5$, and we can thus evaluate both with a value of $\tr_5$ such that $\text{Im}\left[\tr_5\right]>0$ ---~as by default in \textsc{PentagonFunctions++}~--- regardless of the permutations or space-time parity. We must only keep track of the sign of $\tr_5$ in the rational coefficients (see Eqs.~\eqref{eq:finrem_ratcoeffs}), which enters our final expressions for the five-particle finite remainders through the values of the momentum twistors, and is determined by the values of the external momenta through its definition~\eqref{eq:tr5}.
The same holds for the other square roots in the problem, $\sqrt{\Delta_3^{(i)}}$, which appear only in the special function monomials. The polynomials $\Delta_3^{(i)}$ are positive in the physical scattering regions. We adopt the convention of Ref.~\cite{Chicherin:2021dyp} that their square roots are positive, $\sqrt{\Delta_3^{(i)}}>0$, as done in \textsc{PentagonFunctions++}.

In conclusion, we reconstruct the analytic expressions for the minimal set of independent finite remainders, and generate the values of the remaining ones by permuting the former at the numerical evaluation stage. We do this by evaluating the rational coefficients at permuted points, whereas we obtain the values of all permutations of the special functions from the values of the functions at the original phase-space point only. This allows us to minimise the amount of analytic data, whose size may otherwise become problematic, and at the same time evaluate the results efficiently.

\section{Validation}
\label{sec:validation}

In this section we discuss a number of validations performed on the analytic results derived in this work.
First, let us remind the readers that the quantities that we reconstructed analytically are the $L$-loop finite remainders, where the UV and IR poles contained in the $L$-loop bare amplitudes are cancelled by the pole terms according to Eq.~\eqref{eq:finiteremainder}. These pole cancellations already provide a strong consistency check of our calculation. In the following subsections we present further checks.

\subsection{Comparison against full six-point computation}

In order to verify the analytic expressions obtained by detaching the leptonic decay current as described in Section~\ref{sec:reduction}, 
we cross-check them against the helicity amplitudes obtained by computing the six-point process directly using a framework that has been applied to the computation of several 
two-loop amplitudes~\cite{Hartanto:2019uvl,Badger:2021owl,Badger:2021imn,Badger:2021ega}. We perform the full six-point computation numerically using the momentum twistor parameterisation~\eqref{eq:mtvardefs} by assigning rational values to the variables $x_1, \ldots, x_8$ in the rational coefficients and treating the special functions symbolically. We derive numerical results for all the sub-amplitudes ---~$A^{(L)}_{6,u}$, $A^{(L)}_{6,d}$, $A^{(L)}_{6,W}$ and $A^{(L)}_{6,e}$~--- in all four contributing helicity configurations. We find full numerical agreement between the two approaches. 
This provides a further robust check of our analytic computation, where we derived analytic expressions only for the independent helicity configurations and obtained the remaining ones by complex conjugation and permutation of the external momenta.

\subsection{Gauge invariance}

The gauge-invariance structure of the $\wpaj$ amplitude is slightly complicated by the different sources of photon emission, as discussed in Section~\ref{sec:amp}.
The individual sub-amplitudes ($A^{(L)}_{6,i}$ with $i=u,d,W,e$) are not separately gauge invariant in the electroweak (EW) sector. Only linear combinations of them, defined in 
Eq.~\eqref{eq:gaugeinvariantamps}, are. We rewrite them here for convenience,
\begin{equation} \label{eq:gaugeinvcombinations}
\left\{ A^{(L)}_{6,u} + \frac{1}{s_{156}-s_{56}} A^{(L)}_{6,W} \,, \quad 
 A^{(L)}_{6,d} - \frac{1}{s_{156}-s_{56}} A^{(L)}_{6,W} \,, \quad 
 A^{(L)}_{6,e} - \frac{1}{s_{156}-s_{56}} A^{(L)}_{6,W} \right\} \,.
\end{equation}
We verify explicitly that these combinations satisfy the EW Ward identity by replacing the photon polarisation vector with its momentum ($\vareps(p_1) \to p_1$) and checking that the resulting expressions vanish.

The QCD Ward identity (performed by replacing the gluon polarisation vector with its momentum, $\vareps(p_3) \to p_3$), instead, is already satisfied by the individual sub-amplitudes. We checked this explicitly as well.

We further demonstrate the gauge invariance by evaluating the helicity amplitudes using two different sets of reference momenta for the photon and gluon polarisation vectors ($q_1$ and $q_3$), finding perfect agreement.

\subsection{Renormalisation scale dependence}
\label{sec:scaledep}

The $L$-loop finite remainders depend on the renormalisation scale, $\mu$. In deriving analytic results for the process $\wpaj$ we set the renormalisation scale to unity ($\mu=1$).
The dependence on the renormalisation scale of the finite remainders can be restored as follows,
\begin{equation}
F^{(L),i}_6\left(\mu^2\right) = F^{(L),i}_6\left(\mu^2 = 1\right) + \delta F^{(L),i}_6 \left(\mu^2\right) \,,
\label{eq:FiniteRemainderMu}
\end{equation}
where we omitted the dependence on the external momenta to simplify the notation. The $\mu$-restoring terms $\delta F_6^{(L),i}$ are built out of the lower-loop finite remainders evaluated at $\mu^2=1$ and logarithms of $\mu^2$. Explicitly, they are given by
\begingroup
\allowdisplaybreaks
\begin{align}
& \delta F^{(1),1}_6\left(\mu^2\right) = \frac{11}{6} A_6^{(0)} \log\left(\mu^2\right) \,, \\
& \delta F^{(1),n_f}_6\left(\mu^2\right) = - \frac{1}{3} A_6^{(0)} \log\left(\mu^2\right) \,, \\
& \delta F^{(2),1}_6\left(\mu^2\right) = \log(\mu^2) \bigg\lbrace \bigg(\frac{1813}{216} - \frac{11}{36}\pi^2 + 8 \zeta_3\bigg) A^{(0)}_6 + \frac{11}{2}F^{(1),1}_6(1) \bigg\rbrace
                             + \frac{121}{24} A^{(0)}_6 \log^2(\mu^2) \,, \\
& \delta F^{(2),n_f}_6\left(\mu^2\right) = \log(\mu^2) \bigg\lbrace \bigg(\frac{\pi^2}{18} -\frac{77}{18}  \bigg) A^{(0)}_6 
                                                                 - F^{(1),1}_6(1)  + \frac{11}{2}F^{(1),n_f}_6(1) \bigg\rbrace - \frac{11}{6} A^{(0)}_6 \log^2(\mu^2) \,, \\
& \delta F^{(2),n_f^2}_6\left(\mu^2\right) = \log(\mu^2) \bigg\lbrace \frac{10}{27} A^{(0)}_6 -  F^{(1),n_f}_6(1) \bigg\rbrace + \frac{1}{6} A^{(0)}_6 \log^2(\mu^2) \,.
\end{align}
\endgroup
The dependence on the external momenta is understood.
We can then use Eq.~\eqref{eq:FiniteRemainderMu} to check that the finite remainders we computed have the correct scale dependence.
We do this by evaluating the finite remainders at two phase-space points that are connected by a rescaling by some positive factor $a$,
\begin{equation}
\begin{aligned}
& \vec{p} = (p_1,p_2,p_3,p_4,p_5,p_6) \,,  \\ 
& \vec{p}^{\,\prime} = a \, \vec{p} =  (a \, p_1,a \, p_2,a \, p_3,a \, p_4,a \, p_5,a \, p_6)\,.  
\end{aligned}
\end{equation}
These two evaluations allow us to confirm numerically that the finite remainders exhibit the correct scaling behaviour, in the form
\begin{equation}
  \frac{F_6^{(L),i}\left(1,a \, \vec{p}\right)+\delta F_6^{(L),i}\left(a^2,a \, \vec{p} \right)}{A_6^{(0)}\left(a\,\vec{p}\right)} 
= \frac{F_6^{(L),i}\left(1,\vec{p}\right)}{A_6^{(0)}\left(\vec{p}\right)} \,,
\label{eq:scaledepcheck}
\end{equation}
where we have made the dependence on the kinematic point explicit.

\subsection{Tree-level and one-loop checks}

We validated the tree-level and one-loop amplitudes derived in this paper against the results available in the literature.
For the tree-level amplitude we compared our helicity amplitudes against the analytic results presented in Ref.~\cite{Campbell:2021mlr} and additionally, for the full colour tree-level squared matrix elements, against \textsc{Madgraph5}~\cite{Alwall:2014hca} for both processes $W^+\gamma j$ and $W^-\gamma j$.
As for the one-loop amplitudes, we compared our results against the leading colour contributions of the $W^+\gamma j$ amplitudes presented in Ref.~\cite{Campbell:2021mlr}.
In all cases we find perfect agreement.
We would like to point out that our choice of reference vectors for the photon and the gluon is different from the one used in Ref.~\cite{Campbell:2021mlr}. For this reason we compared the gauge invariant combinations of sub-amplitudes shown in Eq.~\eqref{eq:gaugeinvcombinations}. This check therefore further validates the gauge invariance of our result.

\subsection{Four-point amplitude comparison}

We performed a cross check of the four-point amplitudes $A_4^{(L)\mu}$ which contribute to the sub-amplitudes $A^{(L)}_{6,W}$ and $A^{(L)}_{6,e}$ against the results provided in Ref.~\cite{Gehrmann:2011ab} for the scattering process $q\bar{q}\to Vg$. In Ref.~\cite{Gehrmann:2011ab} analytic results are presented 
for the helicity coefficients which are linear combinations of the form factors $b_i^{(L)}$ in Eq.~\eqref{eq:B4decomposition}, evaluated at $\mu^2 = s_{234}$.
In order to enable a direct comparison for the one- and two-loop leading colour finite remainders, we 
recomputed the $A_4^{(L)\mu}$ amplitudes in Eq.~\eqref{eq:B4decomposition} using the tensor structures employed in Ref.~\cite{Gehrmann:2011ab}.
Since we compute the finite remainders with $\mu^2=1$, we obtain the results at $\mu^2=s_{234}$ using the formulae to restore the dependence on $\mu$ shown in Section~\ref{sec:scaledep}. We obtain perfect numerical agreement for the helicity coefficients. We further check that the four-particle finite remainders $F_4^{(L)\mu}$ computed using the tensor structures of Ref.~\cite{Gehrmann:2011ab} match the ones we derived using the tensor structures defined in Section~\ref{sec:B4amplitude} after contracting them with the decay currents according to Eq.~\eqref{eq:decayBF}.

\section{Results}
\label{sec:results}

We provide analytic expressions for the five- and four-point contracted amplitudes ($\tilde{A}^{(L)}_{5,i}$ and $\tilde{A}^{(L)}_{4,i}$), at one and two loops, together with the decay currents ($L_{A,\mu}$, $L^{e}_{B,\mu}$, $L^{W}_{B,\mu}$) and the relevant projection matrices ($\Delta$~\eqref{eq:Delta} and $\Omega$~\eqref{eq:Omega}) 
in the ancillary files. The amplitudes are presented as linear combinations of independent rational coefficients that multiply a monomial basis of square roots and special functions. 

We confirm the previous observations about the cancellation of the pentagon functions involving certain letters~\cite{Badger:2021nhg,Badger:2021ega,Abreu:2021asb}. We observe that the functions involving the letters $\mathcal{Z}=\{W_{18}, W_{25}, W_{34}, W_{45}, W_{46}, W_{57} \}$ (in the notation of Refs.~\cite{Abreu:2021smk,Chicherin:2021dyp}) are present in the contributing integrals but drop out from the amplitudes truncated at order $\eps^0$,\footnote{This holds for the set of independent amplitudes we reconstructed explicitly, which receive contributions only from the cyclic permutations of the master integrals. Since the set of letters $\mathcal{Z}$ is not closed under all $S_4$ permutations, these letters are present in some of the permuted amplitudes which contribute to the helicity and colour summed squared finite remainders.} and that the functions involving the letter $W_{198}=\tr_5$ are present in the amplitudes and drop out from the finite remainders. The latter phenomenon is by now well established in the case of fully massless scattering, where it has been linked to an underlying cluster algebra structure of the letter alphabet~\cite{Chicherin:2020umh}.

We include a \textsc{Mathematica} script to demonstrate the assembly of both the $\wpaj$ and $\wmaj$ amplitudes, and to perform the numerical evaluation of the finite remainders at a given kinematic point.
We evaluate the special functions in the physical scattering region using the package \textsc{PentagonFunctions++}~\cite{Chicherin:2021dyp}, as discussed in Section~\ref{sec:Permutations}.

We use the following configuration of momenta,
\begin{equation}
-p_2 - p_4 \to p_1 + p_3 + p_5 + p_6 \,,
\label{eq:momconfig}
\end{equation}
to define the six scattering channels for $pp\to W^+\gamma j$ production,
\begin{equation}
\label{eq:wplus_channel_definition}
\begin{aligned}
&\mathbf{u\bar{d}}: \quad  u(-p_2) + \bar{d}(-p_4) \to \gamma(p_1) + g(p_3) + \nu_e(p_5) + e^+(p_6) \,, \\
&\mathbf{\bar{d}u}: \quad  \bar{d}(-p_2) + u(-p_4) \to \gamma(p_1) + g(p_3) + \nu_e(p_5) + e^+(p_6) \,, \\
&\mathbf{ug}:       \quad  u(-p_2) + g(-p_4)       \to \gamma(p_1) + d(p_3) + \nu_e(p_5) + e^+(p_6) \,, \\
&\mathbf{gu}:       \quad  g(-p_2) + u(-p_4)       \to \gamma(p_1) + d(p_3) + \nu_e(p_5) + e^+(p_6) \,, \\
&\mathbf{\bar{d}g}: \quad  \bar{d}(-p_2) + g(-p_4) \to \gamma(p_1) + \bar{u}(p_3) + \nu_e(p_5) + e^+(p_6) \,, \\
&\mathbf{g\bar{d}}: \quad  g(-p_2) + \bar{d}(-p_4) \to \gamma(p_1) + \bar{u}(p_3) + \nu_e(p_5) + e^+(p_6) \,,
\end{aligned}
\end{equation}
and similarly for $pp\to W^-\gamma j$ production,
\begin{equation}
\label{eq:wmin_channel_definition}
\begin{aligned}
&\mathbf{d\bar{u}}: \quad  d(-p_2) + \bar{u}(-p_4) \to \gamma(p_1) + g(p_3)       + e^-(p_5) + \bar\nu_e(p_6) \,, \\
&\mathbf{\bar{u}d}: \quad  \bar{u}(-p_2) + d(-p_4) \to \gamma(p_1) + g(p_3)       + e^-(p_5) + \bar\nu_e(p_6) \,, \\
&\mathbf{dg}:       \quad  d(-p_2) + g(-p_4)       \to \gamma(p_1) + u(p_3)       + e^-(p_5) + \bar\nu_e(p_6) \,, \\
&\mathbf{gd}:       \quad  g(-p_2) + d(-p_4)       \to \gamma(p_1) + u(p_3)       + e^-(p_5) + \bar\nu_e(p_6) \,, \\
&\mathbf{\bar{u}g}: \quad  \bar{u}(-p_2) + g(-p_4) \to \gamma(p_1) + \bar{d}(p_3) + e^-(p_5) + \bar\nu_e(p_6) \,, \\
&\mathbf{g\bar{u}}: \quad  g(-p_2) + \bar{u}(-p_4) \to \gamma(p_1) + \bar{d}(p_3) + e^-(p_5) + \bar\nu_e(p_6) \,.
\end{aligned}
\end{equation}
The interference between the $L$-loop finite remainders and the tree-level amplitudes summed over
colour and helicity in the leading colour approximation is given by
\begin{equation}
\sum_{\mathrm{colour}}\sum_{\mathrm{helicity}} \cA_6^{(0)*} \cF_6^{(L)} =: 2 e^2 g_W^4 g_s^2 n^L N_c^2 \, \cH^{(L)} \,,
\end{equation}
where the \textit{reduced squared finite remainder} $\cH^{(L)}$ is defined by
\begin{equation}
\cH^{(L)} = \sum_{\mathrm{helicity}}  A_{6}^{(0)*}  F_{6}^{(L)} \,,
\end{equation}
for all scattering channels given in Eqs.~\eqref{eq:wplus_channel_definition} and~\eqref{eq:wmin_channel_definition}. 
The reduced squared finite remainder obeys the same decomposition according to the closed fermion loop contributions as $F_{6}^{(L)}$,
\begin{align} \label{eq:NfDecompositionH} 
\begin{aligned}
& \cH^{(1)} = N_c \cH^{(1),1} + n_f \cH^{(1),n_f} \,, \\
& \cH^{(2)} = N_c^2 \cH^{(2),1} + N_c n_f \cH^{(2),n_f} + n_f^2 \cH^{(2),n_f^2} \,.
\end{aligned}
\end{align}

We present a benchmark evaluation at the following phase-space point in the physical scattering region specified by Eq.~\ref{eq:momconfig} (the momenta are given in units of GeV),
\begin{align}
\label{eq:PSpoint}
\begin{aligned}
p_1 & = (88.551333054, -22.100690287, 40.080353191, -75.805430956) \,, \\
p_2 & = (-500,0,0,-500) \,,  \\
p_4 & = (328.32941922, -103.84961188, -301.93375538, 76.494921387) \,, \\
p_2 & = (-500,0,0, 500) \,,  \\
p_5 & = (152.35810946, -105.88095966, -97.709638326, 49.548385226)  \,, \\
p_6 & = (430.76113825, 231.83126183, 359.56304052, -50.237875657)   \,, 
\end{aligned}
\end{align}
with $\tr_5 =  2.167055i \cdot 10^{10} \; \mathrm{GeV}^4$. We take the $W$ boson mass and width to be 
\begin{equation}
M_W = 80.4109 \; \mathrm{GeV} \,, \qquad \qquad \Gamma_W = 2.0467 \; \mathrm{GeV}\,.
\label{eq:inputparameters}
\end{equation}
High precision values for the phase space point in Eq.~\eqref{eq:PSpoint} as well as the input parameters in Eq.~\eqref{eq:inputparameters} are provided in the ancillary files.
We present in Tables~\ref{tab:benchmark2Lnf0bare} and~\ref{tab:benchmark2Lnf1bare} the values of the bare two-loop amplitudes
normalised to the tree-level amplitudes in the $\mathbf{u\bar{d}}$ scattering channel for each individual sub-amplitude,
\begin{equation}  \label{eq:treenorm}
\hat{A}_{6,i}^{(L),j} = \frac{A_{6,i}^{(L),j}}{A_{6,i}^{(0)}} \,, 
\end{equation} 
for $i = u,d,W,e$ and the two closed fermion loop contributions specified in Eq.~\eqref{eq:NFdecomposition}, namely $j=1,n_f$.
The results are presented only for the two independent helicity configurations ($\scriptstyle +++--+$ and $\scriptstyle -++--+$).
In Table~\ref{tab:benchmarkfinremsq2L} we show the values of the two-loop reduced squared finite remainders normalised to the reduced squared tree-level amplitudes, 
\begin{equation}
\hat{\cH}^{(L)} = \frac{\cH^{(L)}}{\cH^{(0)}} \,,
\end{equation}
for all channels of both $pp \to \wpaj$ and $pp\to\wmaj$ production. We give analogous tables for the one-loop amplitudes in Appendix~\ref{app:oneloop}.
\begin{table}[t!]
\centering
\begin{tabularx}{1.0\textwidth}{|C{0.7}|C{1.0}|C{0.6}|C{1.0}|C{1.2}|C{1.2}|C{1.3}|}
\hline
      & helicity & $\eps^{-4}$ & $\eps^{-3}$ & $\eps^{-2}$ & $\eps^{-1}$ & $\eps^{0}$ \\
\hline
$\hat A^{(2),1}_{6,u}$ & $\scriptstyle +++--+$ & 2 & -49.5288 & $603.232 + 4.18740 i$ & $-4813.11 - 82.3401 i$ & $28289.7 + 713.980 i$ \\
                   & $\scriptstyle -++--+$ & 2 & -49.5288 & $605.560 + 1.03233 i$ & $-4867.68 - 10.1740 i$ & $28904.1 - 84.4212 i$ \\
\hline
$\hat A^{(2),1}_{6,d}$ & $\scriptstyle +++--+$ & 2 & -49.5288 & $606.017 + 4.37613 i$ & $-4883.27 - 87.6955 i$ & $29148.2 + 787.284 i$ \\
                   & $\scriptstyle -++--+$ & 2 & -49.5288 & $604.589 + 4.36093 i$ & $-4848.83 - 90.4281 i$ & $28743.3 + 856.481 i$ \\
\hline
$\hat A^{(2),1}_{6,W}$ & $\scriptstyle +++--+$ & 2 & -49.5288 & $605.100 + 3.07126 i$ & $-4859.29 - 58.7793 i$ & $28844.2 + 480.026 i$ \\
                   & $\scriptstyle -++--+$ & 2 & -49.5288 & $605.637 + 2.40762 i$ & $-4871.59 - 43.1992 i$ & $28978.3 + 302.671 i$ \\
\hline
$\hat A^{(2),1}_{6,e}$ & $\scriptstyle +++--+$ & 2 & -49.5288 & $605.140 + 2.93702 i$ & $-4860.19 - 55.6437 i$ & $28853.6 + 444.669 i$ \\
                   & $\scriptstyle -++--+$ & 2 & -49.5288 & $606.606 + 2.97710 i$ & $-4894.35 - 56.2398 i$ & $29236.9 + 444.300 i$ \\
\hline
\end{tabularx}
\caption{\label{tab:benchmark2Lnf0bare} 
Bare two-loop helicity sub-amplitudes (normalised to the tree-level amplitudes as in Eq.~\eqref{eq:treenorm}) without any closed fermion loop contribution 
for $\wpaj$ production in the $\mathbf{u\bar{d}}$ scattering channel 
evaluated at the kinematic point given in Eq.~\eqref{eq:PSpoint}. The results are shown for the two independent helicity configurations and obtained with $q_1 = p_3$ and $q_3 = p_1$ where $q_1$ ($q_3$) is the 
reference momentum for the photon (gluon) polarisation vector.
}
\end{table}
\begin{table}[t!]
\centering
\begin{tabularx}{1.0\textwidth}{|C{0.7}|C{1.0}|C{0.6}|C{1.0}|C{1.2}|C{1.2}|C{1.3}|}
\hline
      & helicity & $\eps^{-4}$ & $\eps^{-3}$ & $\eps^{-2}$ & $\eps^{-1}$ & $\eps^{0}$ \\
\hline
$\hat A^{(2),n_f}_{6,u}$ & $\scriptstyle +++--+$ & 0 & 0.333333 & -7.39369 & $79.8302 + 1.39580 i $ & $-556.215 - 14.3791 i$ \\ 
                     & $\scriptstyle -++--+$ & 0 & 0.333333 & -7.39369 & $80.6063 + 0.34411 i $ & $-570.821 + 1.89741 i$ \\
\hline
$\hat A^{(2),n_f}_{6,d}$ & $\scriptstyle +++--+$ & 0 & 0.333333 & -7.39369 & $80.7586 + 1.45871 i $ & $-576.048 - 17.8798 i$ \\
                     & $\scriptstyle -++--+$ & 0 & 0.333333 & -7.39369 & $80.2827 + 1.45364 i $ & $-566.721 - 19.4318 i$ \\
\hline
$\hat A^{(2),n_f}_{6,W}$ & $\scriptstyle +++--+$ & 0 & 0.333333 & -7.39369 & $80.4531 + 1.02375 i $ & $-569.113 - 9.88866 i$ \\
                     & $\scriptstyle -++--+$ & 0 & 0.333333 & -7.39369 & $80.6321 + 0.802539 i$ & $-572.278 - 6.31448 i$ \\
\hline
$\hat A^{(2),n_f}_{6,e}$ & $\scriptstyle +++--+$ & 0 & 0.333333 & -7.39369 & $80.4664 + 0.979007 i$ & $-569.368 - 9.14880 i$ \\
                     & $\scriptstyle -++--+$ & 0 & 0.333333 & -7.39369 & $80.9551 + 0.992365 i$ & $-577.544 - 9.72557 i$ \\
\hline
\end{tabularx}
\caption{\label{tab:benchmark2Lnf1bare} 
Bare two-loop helicity sub-amplitudes (normalised to the tree-level amplitudes as in Eq.~\eqref{eq:treenorm}) with one closed fermion loop  
for $\wpaj$ production in the $\mathbf{u\bar{d}}$ scattering channel 
evaluated at the kinematic point given in Eq.~\eqref{eq:PSpoint}. The results are shown for the two independent helicity configurations and obtained with $q_1 = p_3$ and $q_3 = p_1$ where $q_1$ ($q_3$) is the 
reference momentum for the photon (gluon) polarisation vector.
}
\end{table}
\begin{table}[t!]
\centering
\begin{tabularx}{1.0\textwidth}{|C{0.4}|C{1.2}|C{1.2}|C{1.2}|}
\hline
$W^+ \gamma j $ & $\mathrm{Re}\,\hat{\mathcal{H}}^{(2),1}$ & $\mathrm{Re}\,\hat{\mathcal{H}}^{(2),n_f}$ & $\mathrm{Re}\,\hat{\mathcal{H}}^{(2),n_f^2}$ \\
\hline
$\mathbf{u\bar{d}}$  & 483.506205134 & -222.568846475 & 22.1747738519  \\
$\mathbf{\bar{d}u}$  & 462.732386147 & -219.389809502 & 22.1747738519  \\
$\mathbf{ug}      $  & 894.669569294 & -309.802310098 & 24.2425489305  \\
$\mathbf{gu}      $  & 796.031872994 & -288.292629199 & 23.3127252902  \\
$\mathbf{\bar{d}g}$  & 954.097242371 & -317.336400774 & 24.2425489305  \\
$\mathbf{g\bar{d}}$  & 898.961273740 & -302.856612446 & 23.3127252902  \\
\hline
\hline
$W^- \gamma j $ & $\mathrm{Re}\,\hat{\mathcal{H}}^{(2),1}$ & $\mathrm{Re}\,\hat{\mathcal{H}}^{(2),n_f}$ & $\mathrm{Re}\,\hat{\mathcal{H}}^{(2),n_f^2}$ \\
\hline
$\mathbf{d\bar{u}}$  & 498.332524932 & -222.702160434 & 22.1747738519  \\ 
$\mathbf{\bar{u}d}$  & 732.600496818 & -268.121335492 & 22.1747738519  \\
$\mathbf{dg}      $  & 1786.14253164 & -305.863467669 & 24.2425489305  \\
$\mathbf{gd}      $  & 1612.34790163 & -407.732735568 & 23.3127252902  \\
$\mathbf{\bar{u}g}$  & 320.710353060 & -152.382317276 & 24.2425489305  \\
$\mathbf{g\bar{u}}$  & 1300.37372328 & -375.944229843 & 23.3127252902  \\
\hline
\end{tabularx}
\caption{\label{tab:benchmarkfinremsq2L} 
Reduced squared finite remainders (normalised to the reduced squared tree level amplitudes) for all closed fermion loop contributions and scattering channels
evaluated at the kinematic point given in Eq.~\eqref{eq:PSpoint} for both $pp\to\wpaj$ and $pp\to\wmaj$ production.
}
\end{table}

\begin{figure}[t!]
\centering
\begin{subfigure}{.5\textwidth}
    \centering
    \includegraphics[width=.9\textwidth]{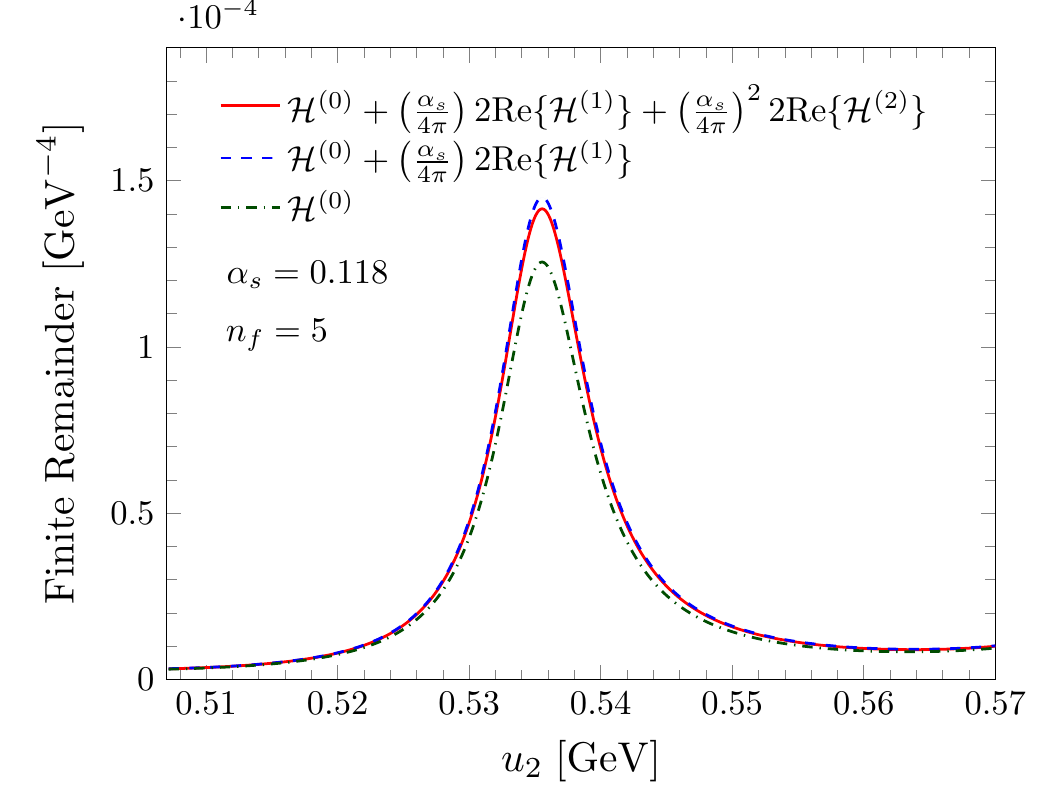}
    \label{fig:uD}
    \caption{$\mathbf{u\bar{d}}$}
\end{subfigure}%
\begin{subfigure}{.5\textwidth}
    \centering
    \includegraphics[width=.9\textwidth]{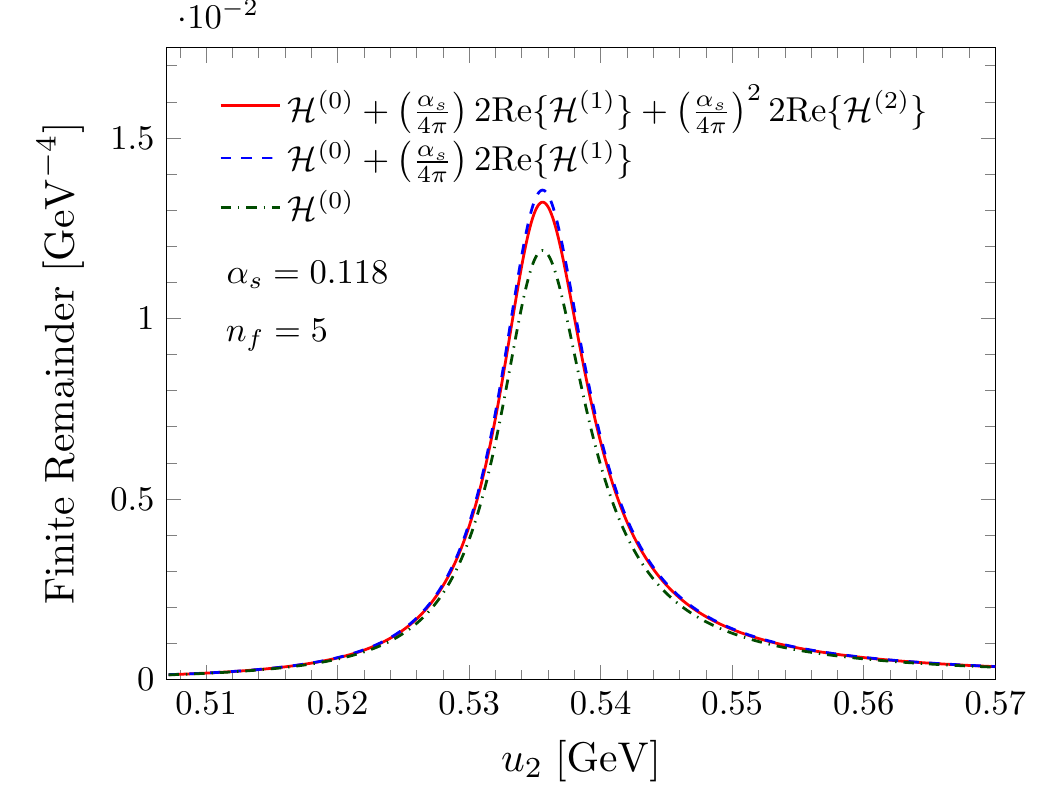}
    \label{fig:Du}
    \caption{$\mathbf{\bar{d}u}$}
\end{subfigure}
\begin{subfigure}{.5\textwidth}
    \centering
    \includegraphics[width=.9\textwidth]{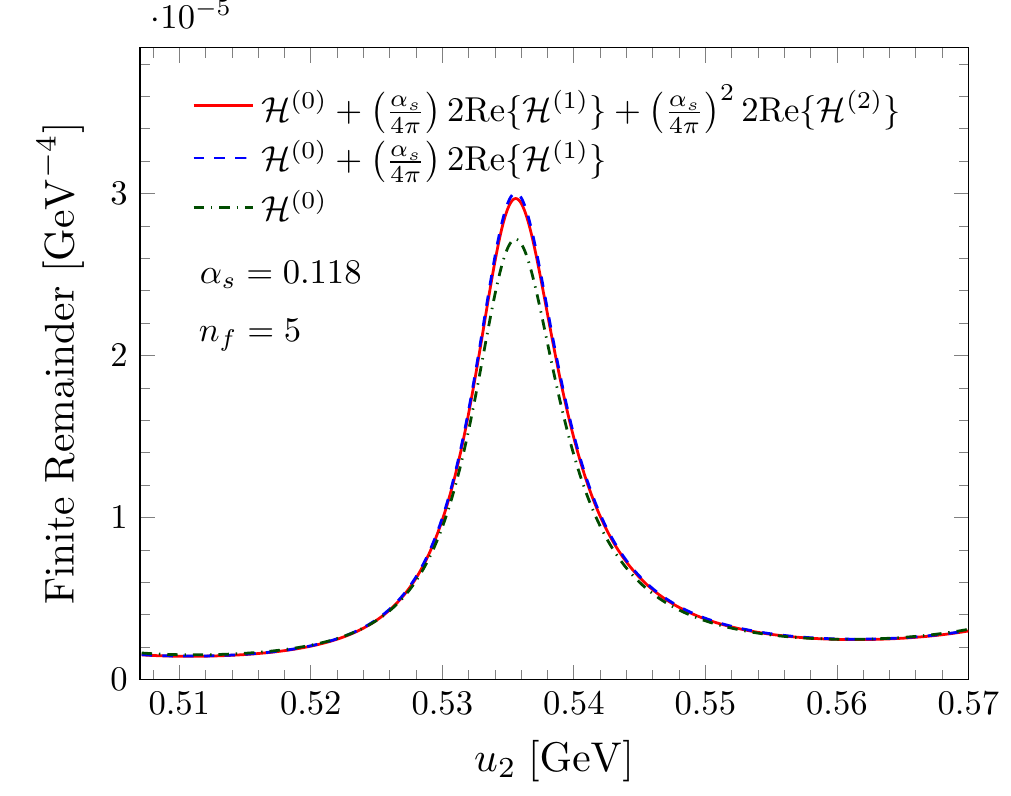}
    \label{fig:ug}
    \caption{$\mathbf{ug}$}
\end{subfigure}%
\begin{subfigure}{.5\textwidth}
    \centering
    \includegraphics[width=.9\textwidth]{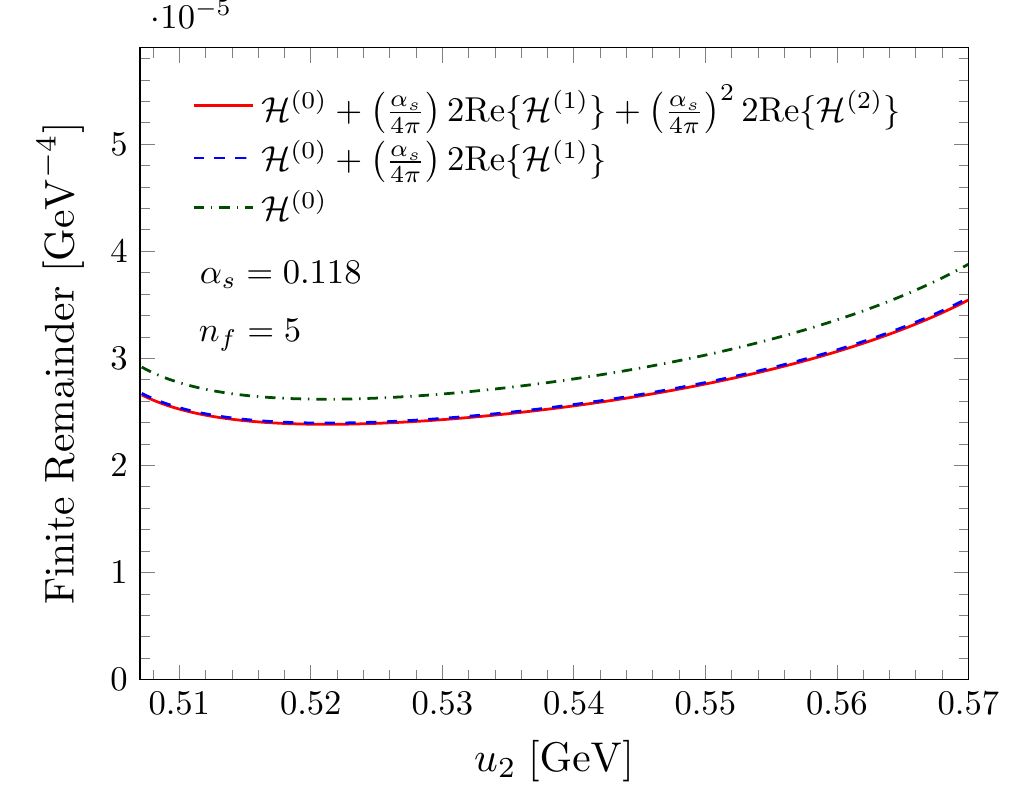}
    \label{fig:gu}
    \caption{$\mathbf{gu}$}
\end{subfigure}
\begin{subfigure}{.5\textwidth}
    \centering
    \includegraphics[width=.9\textwidth]{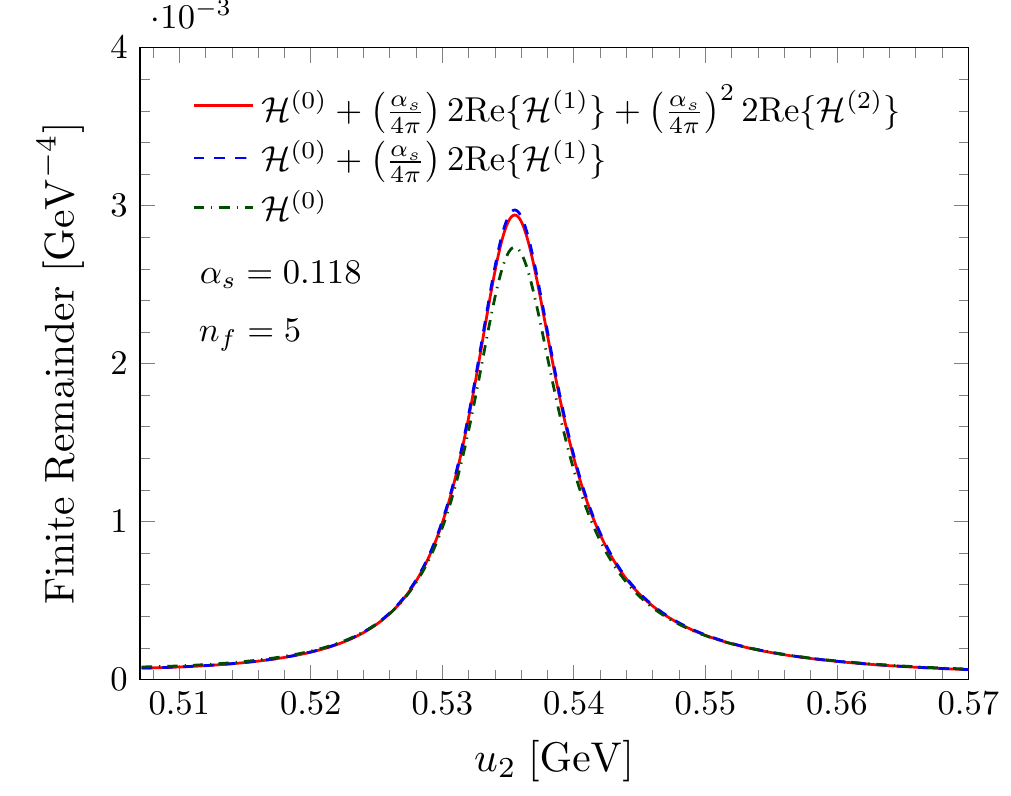}
    \label{fig:Dg}
    \caption{$\mathbf{\bar{d}g}$}
\end{subfigure}%
\begin{subfigure}{.5\textwidth}
    \centering
    \includegraphics[width=.9\textwidth]{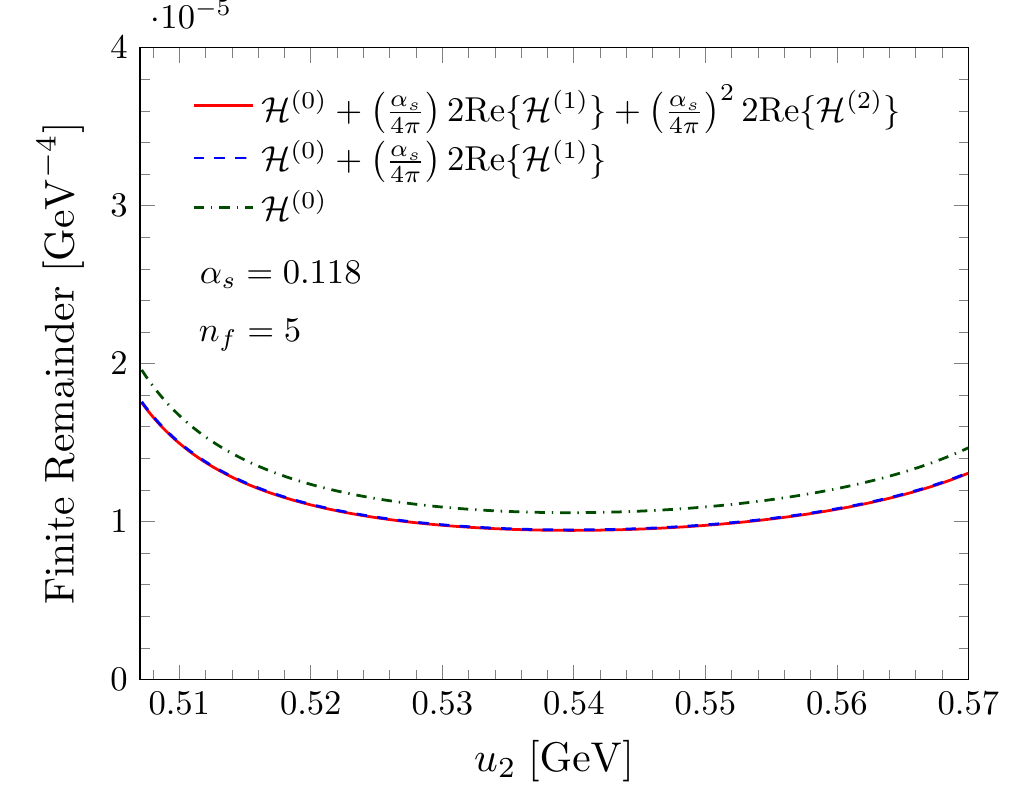}
    \label{fig:gD}
    \caption{$\mathbf{g\bar{d}}$}
\end{subfigure}
\caption[short]{Reduced squared finite remainders $\mathcal{H}^{(L)}$ at tree level, one and two loops evaluated on the univariate phase-space slice defined by Eqs.~\eqref{eq:unislice1}, \eqref{eq:unislice2} and~\eqref{eq:unislice3}, with the parameters given in Eq.~\eqref{eq:unisliceparams}, for all channels of $\wpaj$ production defined in Eq.~\eqref{eq:wplus_channel_definition}.}
\label{fig:plots}
\end{figure}

In order to show the suitability and stability of our evaluation strategy, we present in Figure~\ref{fig:plots} the evaluation of the reduced squared finite remainders on a one-dimensional slice of the physical phase space for all channels of $\wpaj$ production. We begin by parameterising the momenta of the one-mass five-particle process relevant for the $W$-production amplitudes as
\begin{align} \label{eq:unislice1}
\begin{aligned}
& p_1^{\mu} = u_1 \, \frac{\sqrt{s}}{2} \, \left(1,1,0,0 \right) \,, \\
& p_2^{\mu} =  \frac{\sqrt{s}}{2} \, \left(-1,0,0,-1 \right) \,, \\
& p_3^{\mu} = u_2 \, \frac{\sqrt{s}}{2} \, \left(1,\cos\theta,-\sin \phi \sin \theta, -\cos \phi \sin\theta \right) \,, \\
& p_4^{\mu} =  \frac{\sqrt{s}}{2} \, \left(-1,0,0,1 \right) \,. \\
\end{aligned}
\end{align}
We fix the value of $\cos\theta$ by requiring that 
\begin{align} \label{eq:unislice2}
(p_5+p_6)^2 = M^2_{ll} \,, 
\end{align}
where $M^2_{ll}$ is the invariant mass of the leptonic pair.
We then parameterise the momenta of the leptonic pair,
\begin{align} \label{eq:unislice3}
p_5^{\mu} = u_3 \, \frac{\sqrt{s}}{2}  \left(1,\cos\theta_{ll},-\sin \phi_{ll} \sin \theta_{ll}, -\cos \phi_{ll} \sin\theta_{ll} \right) \,,
\end{align}
and $p_6$ follows from momentum conservation.
We fix $u_3$ by requiring that $p_6^2 = 0$. In order to define a univariate phase-space slice, we choose
\begin{align} \label{eq:unisliceparams}
s = 10^4 \, \text{GeV}^2 \,, \qquad M_{ll} = 60 \, \text{GeV}  \,, \qquad \phi = \frac{1}{10} \,, \qquad u_1 = \frac{1}{7}\,, \qquad \theta_{ll} = \frac{\pi}{2} \,, \qquad \phi_{ll} = \frac{\pi}{3} \,.
\end{align}
The remaining variable, $u_2$, is constrained to the interval $[87/175,29/50]$. We chose these values arbitrarily so that the slice crosses a number of spurious poles, i.e.\ points where the rational coefficients diverge whereas the finite remainders stay finite. We checked explicitly that, while approaching such spurious poles, the values of the rational coefficients become larger and larger, while the finite remainders converge. This is a robust check of the stability of the evaluation, since the convergence requires large numerical cancellations among various terms of the finite remainders. 
Figure~\ref{fig:plots} shows the plots of the reduced squared finite remainders up to two-loop order for all channels of $\wpaj$ production on the univariate phase-space slice defined above. 

\section{Conclusions}
\label{sec:conclusions}

In this article we have presented the two-loop leading colour QCD helicity amplitudes for the
process $W^{\pm}\gamma j$ for the first time. We have obtained relatively compact analytic
expressions that can be efficiently evaluated across the full physical phase-space. We constructed
the colour and helicity summed finite remainders, and performed several validation tests. This opens
the path for precision phenomenological predictions at NNLO accuracy in the strong coupling.

To obtain the best possible theoretical predictions it will be necessary to improve upon the leading
colour approximation taken in this article. While it is expected that the leading colour contribution
dominates, a quantitative statement is not possible without explicit computation. Sub-leading colour
corrections require non-planar topologies to be taken into account, and represent a considerable
increase in analytic and algebraic complexity. Progress on the relevant Feynman integrals has been made in this direction quite
recently~\cite{Abreu:2021smk}, although a few topologies contributing to the full amplitudes are
still missing. We note that the missing closed fermion loop contributions, $A^{(2)}_{6,q}$, only
require non-planar hexaboxes and therefore could be considered on a shorter timescale.

We also hope that our approach to the simplification of the reconstructed amplitudes will be of use in
subsequent amplitude computations. An improved understanding of how a rational parameterisation can
be tuned to simplify a particular rational coefficient would certainly be of great value. We expect
this to be of particular importance when dealing with sub-leading colour and non-planar
configurations, in which many different orderings appear simultaneously. It would also be interesting
to study the effect of this method on the reconstruction of the amplitude, i.e.\ whether the
reconstruction is performed in terms of $s_{ij}, \tr_5$ variables or a rational parametrisation.

\acknowledgments
SZ wishes to thank Yang Zhang for useful discussions about partial fraction decomposition. We would
also like to thank Tiziano Peraro for useful discussions and comments on the manuscript. This
project received funding from the European Union's Horizon 2020 research and innovation programmes
\textit{New level of theoretical precision for LHC Run 2 and beyond} (grant agreement No 683211),
and \textit{High precision multi-jet dynamics at the LHC} (grant agreement No 772009).  HBH was
partially supported by STFC consolidated HEP theory grant ST/T000694/1.  SZ gratefully acknowledges
the computing resources provided by the Max Planck Institute for Physics and by the Max Planck
Computing \& Data Facility.

\appendix

\section{Renormalisation Constants}
\label{app:renormconstants}

In this appendix we list the values of the $\beta$ function coefficients and anomalous dimensions relevant for the IR and UV structure of the amplitudes discussed in Section~\ref{sec:amp}~\cite{Becher:2009qa}:
\begingroup
\allowdisplaybreaks
\begin{align}
\beta_0 = & \;\frac{11}{3}C_A - \frac{4}{3} T_F n_f \,, \\
\beta_1 = & \; \frac{34}{3} C_A^2 - \frac{20}{3} C_A T_F n_f - 4 C_F T_F n_f \,, \\
\gamma_0^g = & \; -\frac{11}{3}C_A + \frac{4}{3} T_F n_f \,,   \\
\gamma_1^g = & \;  C_A^2 \left( -\frac{692}{27} + \frac{11\pi^2}{18} + 2 \zeta_3\right)
               + 4 C_F T_F n_f
               + C_A T_F n_f \left( \frac{256}{27} - \frac{2\pi^2}{9}\right) \,, \\
\gamma_0^q = & \; -3 C_F \,, \\
\gamma_1^q = & \; C_F^2 \left( -\frac{3}{2} + 2 \pi^2 - 24 \zeta_3 \right)
                  + C_F C_A \left( -\frac{961}{54} -\frac{11\pi^2}{6} + 26 \zeta_3 \right) \nn
             & \;  + C_F T_F n_f \left( \frac{130}{27} + \frac{2\pi^2}{3} \right) \,, \\
\gamma_0^\cusp = & \; 4 \,, \\
\gamma_1^\cusp = & \; \left( \frac{268}{9} - \frac{4\pi^2}{3} \right) C_A -\frac{80}{9} T_F n_f \,,
\end{align}
\endgroup
where $C_A = N_c$, $C_F = (N_c^2-1)/(2N_c)$, and $T_F= 1/2$.

\section{One-Loop Results}
\label{app:oneloop}

We show in Table~\ref{tab:benchmark1Lnf0bare} the numerical values of the one-loop bare sub-amplitudes normalised to the tree-level amplitudes ($\hat{A}^{(1),1}_{i}$ for $i=u,d,W,e$) in the $\mathbf{u\bar{d}}$ scattering channel evaluated at the kinematic point given in Eq.~\eqref{eq:PSpoint}.
The corresponding reduced squared tree-level amplitudes $\cH^{(0)}$ and reduced squared one-loop finite remainders normalised to the reduced squared tree-level amplitudes ($\hat\cH^{(1),j}$ with $j=1,n_f$) are shown in Table~\ref{tab:benchmarkfinremsq1L} for both $\wpaj$ and $\wmaj$ production.
\begin{table}[t!]
\centering
\begin{tabularx}{1.0\textwidth}{|C{0.7}|C{1.0}|C{0.6}|C{1.0}|C{1.2}|C{1.2}|C{1.3}|}
\hline
      & helicity & $\eps^{-2}$ & $\eps^{-1}$ & $\eps^{0}$ & $\eps^{1}$ & $\eps^{2}$ \\
\hline
$\hat A^{(1),1}_u$ & $\scriptstyle +++--+$ & -2 & 23.8477 & $-138.615 - 2.09370  i$ & $523.949 + 12.3666 i$ & $-1448.23 - 21.7701 i$ \\
                   & $\scriptstyle -++--+$ & -2 & 23.8477 & $-139.779 - 0.516164 i$ & $535.218 - 2.01397 i$ & $-1503.32 + 46.1044 i$ \\
\hline
$\hat A^{(1),1}_d$ & $\scriptstyle +++--+$ & -2 & 23.8477 & $-140.008 - 2.18806 i$  & $539.871 + 13.7461 i$ & $-1538.48 - 30.4669 i$ \\
                   & $\scriptstyle -++--+$ & -2 & 23.8477 & $-139.294 - 2.18046 i$  & $532.471 + 15.2170 i$ & $-1499.51 - 44.3866 i$ \\
\hline
$\hat A^{(1),1}_W$ & $\scriptstyle +++--+$ & -2 & 23.8477 & $-139.550 - 1.53563 i$  & $534.185 + 8.26368 i$ & $-1503.99 - 6.62167 i$ \\
                   & $\scriptstyle -++--+$ & -2 & 23.8477 & $-139.818 - 1.20381 i$  & $536.639 + 5.03856 i$ & $-1515.10 + 9.39205 i$ \\
\hline
$\hat A^{(1),1}_e$ & $\scriptstyle +++--+$ & -2 & 23.8477 & $-139.570 - 1.46851 i$  & $534.360 + 7.61926 i$ & $-1504.71 - 3.48177 i$ \\
                   & $\scriptstyle -++--+$ & -2 & 23.8477 & $-140.303 - 1.48855 i$  & $541.353 + 7.64165 i$ & $-1538.44 - 2.31498 i$ \\
\hline
\end{tabularx}
\caption{\label{tab:benchmark1Lnf0bare} 
Bare one-loop helicity sub-amplitudes (normalised to the tree-level amplitudes as in Eq.~\eqref{eq:treenorm}) without any closed fermion loop contribution  
for $\wpaj$ production in the $\mathbf{u\bar{d}}$ scattering channel 
evaluated at the kinematic point given in Eq.~\eqref{eq:PSpoint}. The results are shown for the two independent helicity configurations and obtained with $q_1 = p_3$ and $q_3 = p_1$ where $q_1$ ($q_3$) is the 
reference momentum for the photon (gluon) polarisation vector.
}
\end{table}

\begin{table}[t!]
\centering
\begin{tabularx}{1\textwidth}{|C{0.4}|C{1.2}|C{1.2}|C{1.2}|}
\hline
$W^+\gamma j$ & $\mathcal{H}^{(0)}  \; [\times 10^{-10}\,\mathrm{GeV}^{-4}] $ & $\mathrm{Re}\;\hat{\mathcal{H}}^{(1),1}$ & $\mathrm{Re}\;\hat{\mathcal{H}}^{(1),n_f}$  \\
\hline
$\mathbf{u\bar{d}}$ & 32.9224527109  & -20.4269208141 & 4.22462265354  \\
$\mathbf{\bar{d}u}$ & 35.8863373066  & -20.0350027848 & 4.22462265354  \\
$\mathbf{ug}      $ & 4.84655650134  & -26.9389515414 & 4.45445318051  \\
$\mathbf{gu}      $ & 15.2151742999  & -25.3235043118 & 4.37533965902  \\
$\mathbf{\bar{d}g}$ & 9.18270882925  & -28.3542876136 & 4.45445318051  \\
$\mathbf{g\bar{d}}$ & 26.4333120479  & -27.3120879601 & 4.37533965902  \\
\hline
\hline
$W^-\gamma j$ & $\mathcal{H}^{(0)}  \; [\times 10^{-10}\,\mathrm{GeV}^{-4}] $ & $\mathrm{Re}\;\hat{\mathcal{H}}^{(1),1}$ & $\mathrm{Re}\;\hat{\mathcal{H}}^{(1),n_f}$  \\
\hline
$\mathbf{d\bar{u}}$ & 48.5521763841  & -20.5759435967 & 4.22462265354  \\
$\mathbf{\bar{u}d}$ & 5.60724308955  & -25.0921274652 & 4.22462265354  \\
$\mathbf{dg}      $ & 0.161819754065 & -53.2745933316 & 4.45445318051  \\
$\mathbf{gd}      $ & 2.59919214772  & -35.7387232774 & 4.37533965902  \\
$\mathbf{\bar{u}g}$ & 0.471356750696 & -25.5067063821 & 4.45445318051  \\
$\mathbf{g\bar{u}}$ & 27.6357549618  & -32.8902240077 & 4.37533965902  \\
\hline
\end{tabularx}
\caption{\label{tab:benchmarkfinremsq1L} 
Reduced squared tree-level amplitude (absolute) and reduced squared one-loop finite remainders (normalised to the reduced squared tree amplitudes)
for the various closed fermion loop contributions and scattering channels of both $pp\to\wpaj$ and $pp\to\wmaj$ production,
evaluated at the kinematic point given in Eq.~\eqref{eq:PSpoint}.
}
\end{table}

\bibliographystyle{JHEP}
\bibliography{planar_waj}

\end{document}